%% file: ISHSMM-arXiv.tex
\documentclass[reqno,oneside,a4paper]{amsart}

\usepackage{graphicx}
\usepackage{caption}
\usepackage{subcaption}
\usepackage{framed}

\usepackage[style=numeric,natbib=true,sortcites=true,firstinits=true,backend=bibtex]{biblatex}

\usepackage{xspace}
\usepackage[usenames,dvipsnames]{color}
\newcommand{\SIHMM}{ISHSMM\xspace} 

\input{my-latex-defs.tex}
\newcommand{\note}[1]{\textbf{\textcolor{blue}{#1}}}

\long\def\redact#1{}
\long\def\othermodels#1{#1}
\long\def\generativemodel#1{}

\newcommand{\GP}{$\Gamma$P\xspace}
\newcommand{\mathGP}{\Gamma \mathrm P}
\newcommand{\bsym}{\boldsymbol}
\newcommand{\DP}{\mathsf{DP}}
\newcommand{\Dir}{\mathsf{Dir}}

\newcommand{\GammaRV}{\mathsf{Gamma}}
\newcommand{\Unif}{\mathsf{Unif}}

\newcommand{\GEM}{\textnormal{GEM}}
\newcommand{\Ind}{\mathbb{I}}

\bibliography{mybib}

\begin{document}

\title{
Infinite Structured Hidden Semi-Markov Models}

\date{\today}

\author[J.~H.~Huggins]{Jonathan H.~Huggins}
\address{Massachusetts Institute of Technology}
\urladdr{http://jhhuggins.org/}
\email{jhuggins@mit.edu}

\author[F.~Wood]{Frank Wood}
\address{University of Oxford}
\urladdr{http://www.robots.ox.ac.uk/~fwood/}
\email{fwood@robots.ox.ac.uk}

%

\begin{abstract}
This paper reviews recent advances in Bayesian nonparametric techniques for constructing and performing inference in infinite hidden Markov models. We focus on variants of Bayesian nonparametric hidden Markov models that enhance a posteriori state-persistence in particular. This paper also introduces a new Bayesian nonparametric framework for generating left-to-right and other structured, explicit-duration infinite hidden Markov models that we call the {\em infinite structured hidden semi-Markov model}.  
\end{abstract}

\maketitle

\section{Introduction}

Parametric hidden Markov models (HMMs), i.e.~those possessing a finite number of states, are now textbook material and widely employed.  Applications of HMMs arise from domains as diverse as speech recognition \cite{Jelinek97}\cite{Juang85}, natural language processing \cite{Manning1999}, hand-writing recognition \cite{Nag1986},  biological sequence modeling \cite{Krogh1994}, gesture recognition \cite{tanguay1995hidden}\cite{wilson1999parametric}, and financial data modeling \cite{ryden1998stylized}.

However, shortcomings of parametric HMMs have long been recognized.  The most commonly identified weakness is the fundamentally combinatorial challenge of determining how many states the HMM should possess.  Another of the most commonly identified shortcomings, highlighted even in Rabiner's seminal tutorial  \cite{Rabiner89}, is the problem of how to bias HMM learning towards models that exhibit long state dwell-times.

Determining out how many states a parametric HMM should have usually involves cross-validation or some other kind of complexity-penalizing model selection procedure.  The Bayesian nonparametric approach replaces this procedure with standard Bayesian inference in a model with an infinite number of states.  While this introduces some mathematical and algorithmic complexity, the overhead is often actually quite small in practice, at least in comparison to Bayesian approaches to HMM learning and inference.  
A number of Bayesian nonparametric HMMs have also appeared in the literature \cite{Beal:2002}\cite{Teh:2006b}\cite{VanGael:2008b}\cite{Heller:2009}; all directly address the problem of learning HMM state cardinality in this way.   The simpler ``infinite HMM'' name for such models applies specifically to the original infinite-state HMM \cite{Beal:2002} although in common usage ``infinite HMM'' is taken to be synonymous to ``Bayesian nonparametric HMM.''  We will use both interchangeably when there is no risk of confusion.   

Enhancing HMM state-dwell-duration has received extensive treatment in the parametric HMM literature \cite{Rabiner89}\cite{Murphy02}\cite{Yu10}\cite{Dewar:2012} where the focus has largely been on the development of algorithms for learning in models where each HMM state is imbued with an explicit dwell-time distribution.   

There are multiple practical reasons for seeking an HMM whose states have this characteristic.  For example some systems' latent states are only distinguishable by dwell-duration (Morse code for example) and as such direct inference about the dwell-duration characteristics may be a requirement.  At least as often, using an HMM to segment natural data results in rapid switching between shorter segments than is desired.  In order to bias segmentation results towards longer steady-state segments requires either building bias into the model towards self-transition or explicitly parameterizing and controlling state dwell-time distributions.  

In the case of infinite HMMs this rapid state-switching problem is exacerbated due to the flexibility that derives from having an infinite number of states. This has led to the development of infinite HMMs that either explicitly bias towards self-transition \cite{Beal:2002}\cite{Fox:2010tg} or explicitly parameterize state-dwell duration \cite{Johnson:2013}.  The infinite structured hidden semi-Markov model (\SIHMM) presented in this paper provides an alternative construction and inference algorithm for such a model.  

The \SIHMM is also a general generative framework for infinite explicit duration HMMs with arbitrary state transition constraints.  This makes it possible to specify, among other things, a Bayesian nonparametric left-to-right HMM.  Left-to-right HMMs are HMMs that never re-visit states.  They are heavily used in practice, particularly for decision-tree like classification and clustering of sequence data.   This nonparametric left-to-right HMM is also a new approach to change point detection in the unknown change point cardinality setting, though this interpretation is largely left for future work.

We introduce the mathematical notation we use in \S\ref{sec:notation}.  \S\ref{sec:review} covers parametric HMMs, including a review of Bayesian approaches to HMM learning, explicit-duration HMMs, and left-to-right HMMs.  
\S\ref{sec:ihmms} reviews infinite HMMs and extensions such as the sticky HDP-HMM. \S\ref{SIHMM} introduces the \SIHMM  generative model and \S\ref{sec:inference} inference algorithms for it. Finally, \S\ref{sec:experiments} explores the \SIHMM experimentally.  

\section{Notation}
\label{sec:notation}
\begin{list}{\labelitemi}{\leftmargin=1em}
\item $[n] := \{ 1,2,3\dots,n \}$ where $n$ is a positive integer. 
\item $\mathbb I(\cdot)$ is the indicator function, which is 1 when the predicate argument is true and 0 otherwise. 
\item $\delta_{\theta}$ is the probability measure concentrated at $\theta$, i.e.~for a set $S$ on which $\delta_{\theta}$ is well-defined, $\delta_{\theta}(S) = \Ind(\theta \in S)$. 
\item $\bsym X$ is a (possibly infinite) matrix, $\bsym x_{i}$ is the $i$-th row of the matrix, and $x_{ij}$ is the  $j$-th component of $\bsym x_{i}$. 
\item $\bsym x$ is a (possibly infinite) vector and $x_{i}$ is the $i$-th component of $\bsym x$.
\item For a vector $\bsym x$, $x_{s:t} = \{ x_{s}, x_{s+1},\dots, x_{t-1}, x_{t} \}$.
\item $\bsym S$ (respectively $\bsym s$) is a matrix (vector) of hidden states.
\item $\bsym Y$ (respectively $\bsym y$) is a matrix (vector) of observed data.
\item $c, \alpha, \alpha_{0}, \alpha_{1}, \dots$ are concentration parameters.
\item $d$ is a discount parameter.
\item Variables of the form $F_{r}, F_{\Theta}, \dots$ are used for data distributions (or measures) while those of the form $H_{r}, H_{\Theta}, \dots$ are used for the corresponding priors over parameters.
\end{list}

\section{Hidden Markov Models}
\label{sec:review}

\begin{figure}[tbp]
\begin{center}
\includegraphics[width=.45\textwidth]{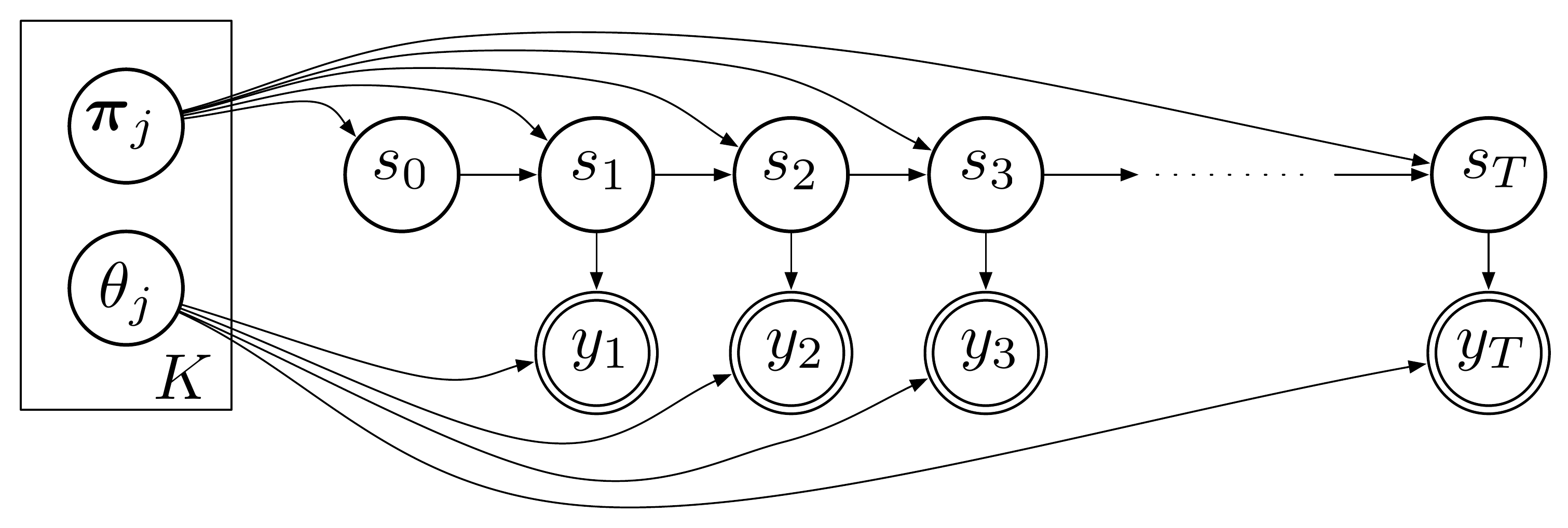}
\caption{HMM Graphical Model}
\label{fig:HMM-graphical-model}
\end{center}
\end{figure}


A hidden Markov model (HMM) consists of a latent Markov process that transitions between a finite number of discrete states at discrete times and is endowed with per state emission distributions.  In the following, $t$ is a discrete index (usually time), $j$ and $i$ are state indices, $\bsym\pi$ is a collection of conditional probability vectors, and $\bsym\pi_{j}$ is a discrete conditional probability distribution:
\eqan
s_{t} | s_{t-1} = i, \bsym\pi &\sim& \mathsf{Discrete}(\bsym\pi_{i}) \label{eq:hmm-gen-state} \\
y_{t} | s_{t} = i, \bsym\theta &\sim& F_{\Theta}(\theta_{i}). \label{eq:hmm-gen-obs}
\enan
The values each element of the latent state $\mathbf s = \{s_1, \ldots, s_T\}$ sequence can take are discrete and finite, i.e.~$s_t \in [K]$. The number $K$ is called the number of states.  The probability of going from state $i$ to state $j$ is given by the $j^{\mathrm{th}}$ entry of the ``transition distribution'' $\bsym\pi_i$, a discrete probability vector whose entries, by definition, sum to one.   The collection of all such transition distributions is $\bsym\pi$.
Each state generates observations from a so-called ``observation'' distribution $F_\Theta$ parameterized by a state-specific parameter $\theta_i \in \Theta$ where $i \in [K].$  


Classical learning in HMMs is textbook machine learning material and is succinctly described as maximum likelihood parameter estimation (the parameters being $\bsym\pi$ and $\bsym\theta$, the collection of all observation distribution parameters)  via expectation maximization where the sum-product algorithm (classically called the forward-backward algorithm) is used to compute expectations over the latent states.   Traditional inference in HMMs often involves finding the most likely latent state trajectory that explains the observed data via the max product algorithm (classically the Viterbi algorithm).  


The number of states in an HMM is sometimes known but usually is not.  Learning becomes substantially more challenging when the number of states must be inferred. This is because some kind of model selection must be performed to choose between HMMs with differing numbers of states. Maximum likelihood learning can be used to determine the number of states only in combination with model complexity penalization such as the Bayesian information criteria or alternatively via cross validation model selection procedures. If a single ``best'' model is the goal for either statistical or computational reasons such approaches can be sufficient.

\subsection{Bayesian HMMs}

An alternative approach to model selection is to specify an HMM with a sufficiently high number of states and regularize the model such that only a small subset of the specified states are actually used.  This is done by placing a prior $H_s$ on $\boldsymbol\pi_j$ that can be parameterized to encourage sparse  $\boldsymbol\pi_j$'s.  A sparse  $\boldsymbol\pi_j$ implies that a small subset of  $\boldsymbol\pi_j$'s entries are non-zero.  This in turn implies that the HMM can transition only to a small subset of states subsequent to state $j$.  If this is true for all states then the total number of HMM states is likely to be small.   The Dirchlet distribution has this characteristic. If $\alpha_0 < 1$, then
\eqan
\boldsymbol\pi_j | H_s &\sim& \Dir(\overbrace{\alpha_{0},\dots,\alpha_{0}}^{K \text{ times}}) \label{eq:hmm-pij} 
\enan
encourages sparsity.  Namely for values of $\alpha_0 \rightarrow 0$ fewer and fewer entries of $\bsym \pi_{j}$ will be non-zero.  It is not difficult to image intuitively how this might work even in the case where the number of states $K\rightarrow \infty$.  
\begin{figure}[tbp]
\begin{center}
\includegraphics[width=.45\textwidth]{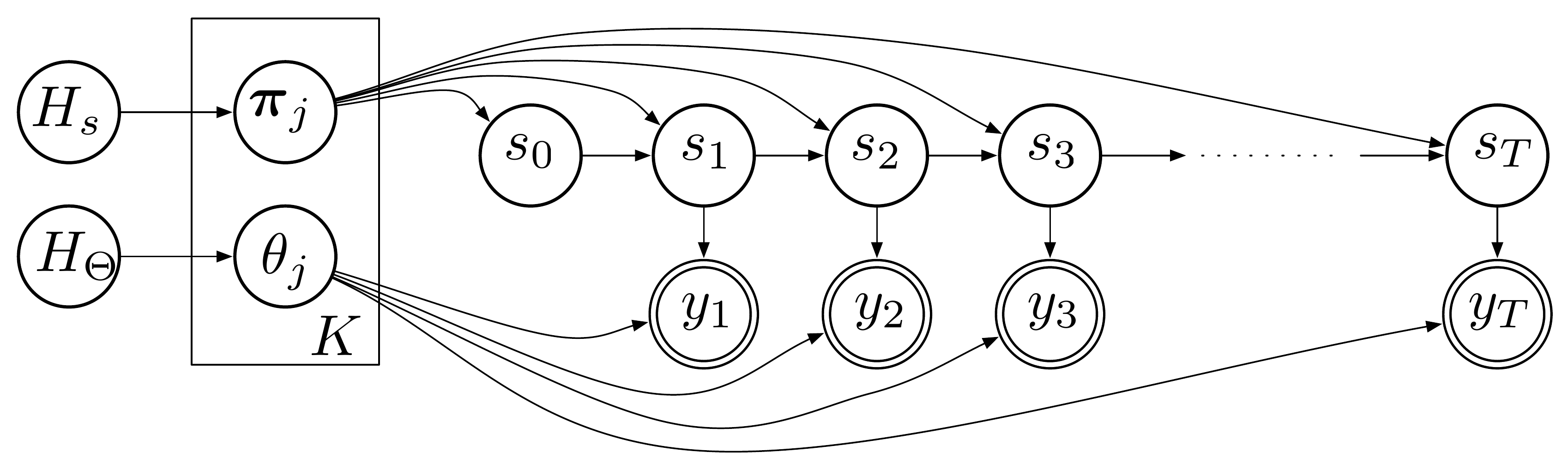}
\caption{Bayesian HMM Graphical Model}
\label{fig:Bayesian-HMM-graphical-model}
\end{center}
\end{figure}
A prior is usually imposed on the observation distribution parameters $\theta_{i} \overset{i.i.d}{\sim} H_{\Theta}$ as well.

When regularized in this way, such an HMM can be interpreted as a Bayesian HMM.  The small graphical change between Figures \ref{fig:HMM-graphical-model} and \ref{fig:Bayesian-HMM-graphical-model} has profound consequences computationally and philosophically beyond simple regularization.  First, in the Bayesian setting the computational learning goal is to estimate a posterior distribution over all latent variables in the model including both the latent state and the observation and transition distribution parameters (in short, a posterior distribution over HMMs).  That is, if we let $\mathcal M$ be the collection of HMM parameters then we can use standard Markov chain Monte Carlo, sequential Monte Carlo, or variational inference techniques to compute an estimate of the posterior $p(\mathcal M | \mathbf Y) \propto p(\mathbf Y | \mathcal M) p(\mathcal M)$.  

This means that when doing, for example, posterior predictive inference, the complexity of the model class used to compute the distribution of the next value of $y_{t+1}$, for instance, is greater than that of a single HMM.  This is because $P(y_{T+1} | \mathbf Y) = \int P( y_{T+1} | \mathcal M) P(\mathcal M | \mathbf Y)d\mathcal M$ is a mixture distribution, each mixture component itself being an HMM.  Another way of thinking about that is that every single sample from the posterior distribution is a different HMM which might give rise to a different segmentation of the observed data, each using a potentially different number of states, each with observation distribution characteristics that may also be different.


\subsubsection{Hierarchical prior for Bayesian HMMs}
\label{sec:hbhmm}
Taking the Bayesian HMM one step further provides insight into the specific Bayesian nonparametric HMMs to come.  In the following hierarchical Bayesian prior,  $\bsym\beta$ is a canonical state transition distribution from which all state specific transition distributions deviate  
\eqan
\bsym\beta | \alpha_{0} &\sim& \Dir(\overbrace{\alpha_{0},\dots,\alpha_{0}}^{K \text{ times}}) \label{eq:hmm-beta} \\
\bsym\pi_{i} | \alpha_{1}, \bsym\beta &\sim& \Dir(\alpha_{1}\bsym \beta). \label{eq:hmm-pi}
\enan
Here the hyperparameter $\alpha_{0}$ controls the overall number of states in the HMM and while $\alpha_{1}$  controls how much the state-specific transition distributions vary from one to another.  

\subsection{Alternate topologies}

In many application settings it is often the case that something is known about the the nature of the latent states.  Often, this comes in the form of information about how transitions between them are restricted.  For example, in a medical diagnostic application that tries to infer a person's chicken-pox state from observable clinical signals, one would want to restrict the latent states such that the state corresponding to pre-chicken-pox could never be reached from the have-had-chicken-pox state.  HMMs with restricted topologies correspond to restricting specific subsets of the entries of the transition distributions to zero.  If an HMM restricts transitions to preclude all states already visited it is called a left-to-right HMM.  Such HMMs are quite common in applications that model processes with hysteresis.  All kinds of topologies can be specified.  In the Bayesian setup these restrictions can be encoded in the prior.  For instance, if a Dirichlet distribution is used as the prior for $\bsym\beta \sim \Dir(\alpha_1, 0, \alpha_3, 0, 0, \ldots, 0, \alpha_K)$, then the resulting HMM will disallow all transitions except to states $1,3,$ and $K$, effectively limiting the complexity of the HMM.  State specific distributions derived from such a sparse $\bsym \beta$ may further restrict the topology of the HMM to preclude, for instance, self transition, or transitions to previously-visited states.

\subsection{Hidden semi-Markov Models }

\begin{figure}[tbp]
\begin{center}
\includegraphics[width=.45\textwidth]{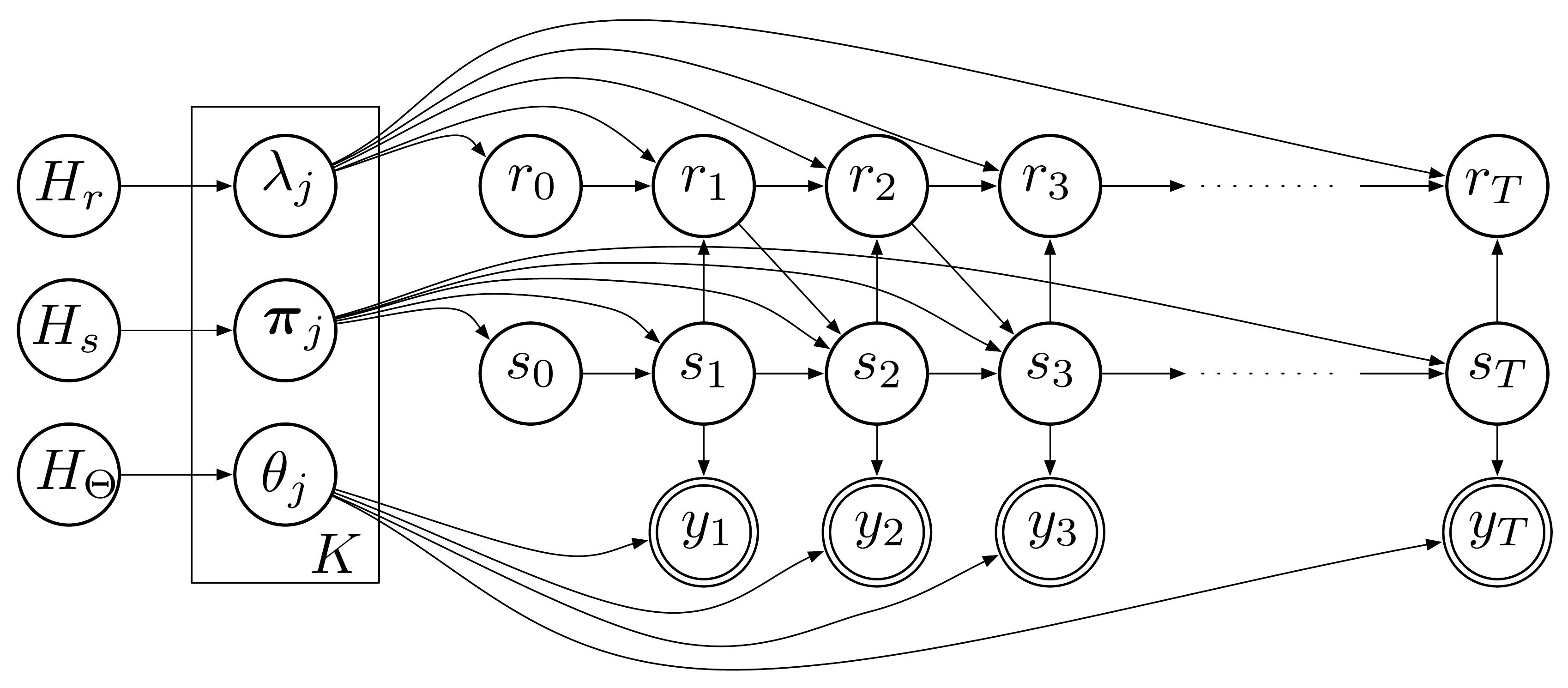}
\caption{Bayesian HSMM Graphical Model}
\label{fig:Bayesian-HSMM-graphical-model}
\end{center}
\end{figure}

One such restricted HMM topology allows no self-transitions.  This is necessary in HMMs that explicitly encode state dwell time.  Such models are known as hidden semi-Markov models (HSMMs) or explicit duration HMMs (EDHMMs).  
Rabiner suggests the need for  HSMMs in his tutorial \cite{Rabiner89} and HSMM applications are common \cite{Gales93}\cite{Ostendorf96}\cite{Zen07}\cite{Yu2006}. 

In such HMMs the latent state is a tuple consisting of a state identifier (as before) and a countdown timer indicating how long the system will remain in that state.  In order to parameterize and control the distribution of state dwell times, self-transition must be disallowed.  Notationally this imposes a second set of latent variables $\mathbf r = \{r_1, \ldots r_T\}$ which are non-negative integers.  The transition dynamic of the latent state consisting of the the tuple with identifier $s_t$ and remaining duration $r_t$ is Markov in the tuple as shown in Fig.~\ref{fig:Bayesian-HSMM-graphical-model}: 
\eqan
s_{t} | s_{t-1} = j, r_{t-1} &\sim&
	\begin{cases} 
	\mathbb I(s_{t} = s_{t-1}), & \; r_{t-1} > 0 \\
	\mathsf{Discrete}(\tilde{\bsym\pi}_{j}), &\; r_{t-1} = 0
	\end{cases}  \label{eq:state-trans-dynamics} \\
r_{t } | s_{t}, r_{t-1} &\sim&
	\begin{cases} 
	\mathbb I(r_{t} = r_{t-1} - 1),& r_{t-1} > 0 \\
	F_{r}(\lambda_{s_{t}}),& r_{t-1} = 0
	\end{cases}  \label{eq:duration-trans-dynamics} 
\enan
Here, the duration counts down until zero when, and only then, the state transitions to a new state.  An example system that has states that are distinguishable only by duration is Morse code.  Dots and dashes have the same observable characteristics, however, they are distinguishable by characteristic distribution.  Modeling such a signal with an HSMM would require three states, two with tone observation distributions and one with background.  The two tone states would be endowed with different duration distribution parameters $\lambda_j$.  Bayesian treatments of HSMM require placing a prior on these parameters as well.  We denote this prior using $H_r$ and write 
$\lambda_j \sim H_{r}.$

\section{Infinite HMMs}
\label{sec:ihmms}

Since we can perform inference over HMMs with varying state cardinality using Bayesian inference tools and an HMM specification with a too-large number of states, it is conceptually straightforward to consider Bayesian inference in HMMs that possess an infinite number of states.  Provided, that is, that priors for transition distributions with an infinite number of bins can be specified and that, additionally, such priors encourage state re-use sufficiently.  Dirichlet processes (DPs) and hierarchical Dirichlet processes (HDPs) have been used to accomplish just these goals.  DPs (infinite analogues of the Dirichlet distribution) allow one to specify a prior over infinite state transition distributions.  Hierarchical compositions of DPs (as in an HDP) allows further encouragement of state reuse. The interested reader is referred to Teh et al. (2006) \cite{Teh:2006b} for a detailed introduction to DPs and HDPs.   

In what follows we review a number of additional infinite HMM constructions. In addition to the standard HDP-HMM, which is the direct Bayesian nonparametric generalization of the finite Bayesian HMM, we highlight two extensions to the HDP-HMM. Both these extensions --- the sticky HDP-HMM and the HDP hidden semi-Markov model --- offer solutions to the state persistence issue.

\subsection{The Hierarchical Dirichlet Process HMM}

The {\em hierarchical Dirichlet process HMM} (HDP-HMM)\cite{Teh:2006b}\cite{VanGael:2008} is the infinite state generalization of the hierarchical Bayesian HMM reviewed in \S\ref{sec:hbhmm}.  It uses the HDP to construct an infinite set of tied conditional transition distributions. 


A two-level HDP for an HDP-HMM can be thought of as an infinite number of transition distribution priors (each a DP), linked together under a top-level DP. The top level DP ensures that the lower level DPs share the same countable set of states and tend to concentrate mass on similar states (referred to as atoms in the DP literature), while still allowing them to have different next-state transition probabilities. 

Such a model, with the top-level DP draw expressed in terms of the stick-breaking construction \cite{Sethuraman:1994}
, is
\eqan
\bsym\beta | \alpha_{0} &\sim& \mathsf{GEM}(\alpha_{0}) \label{eq:hdp-hmm-gem} \\
\bsym\pi_{i} | \bsym\beta, \alpha_{1} &\sim& \textnormal{DP}(\alpha_{1}\bsym\beta) \label{eq:hdp-hmm-pi}  \\
\theta_{i} | H_{\Theta} &\sim& H_{\Theta} \label{eq:hdp-hmm-theta}  \\
s_{t} | s_{t-1} = i, \bsym\pi &\sim& \mathsf{Discrete}(\bsym\pi_{i}) \label{eq:hdp-hmm-state} \\
y_{t} | s_{t} = i, \bsym\theta &\sim& F_{\Theta}(\theta_{i}). \label{eq:hdp-hmm-obs}
\enan
These variables have the same meaning as before, namely, $s$'s are states, $y$'s are observations, $\theta_i$'s are state-specific emission distribution parameters, $\bsym\pi_i$'s are state-specific transition distributions, and $\bsym\beta$ is a global state popularity. For the HDP-HMM, however, $\bsym\beta$ and all $\bsym\pi_i$'s are infinite dimensional. Also, note the similarity between the HDP stick-breaking construction for generating 
$\bsym \beta$ and $\bsym \pi_{i}$ given in \eqref{eq:hdp-hmm-gem} and \eqref{eq:hdp-hmm-pi} and the Bayesian HMM procedure for generating their finite length analogs in \eqref{eq:hmm-beta} and \eqref{eq:hmm-pi}. 

Equations \eqref{eq:hdp-hmm-gem}-\eqref{eq:hdp-hmm-theta} are equivalent to a model which relies on a top-level DP with base measure $H_{\Theta}$
\eqan 
D_{0} | \alpha_{0}, H_{\Theta} &\sim& \DP(\alpha_{0} H_{\Theta}) \label{eq:hdp-top-dp} \\
D_{0} &=& \sum_{j=1}^{\infty} \beta_{j}\delta_{\theta_{j}}
\enan
and where a sequence of draws from a DP with base measure $D_{0}$ are then made, one for each state: 
\eqan
D_{i} | \alpha_{1}, D_{0} &\sim& \DP(\alpha_{1} D_{0})\;\textnormal{for } i = 1,2,3,\dots \label{eq:hdp-lower-dps} 
\\
D_{i} &=& \sum_{j=1}^{\infty} \pi_{ij}\delta_{\theta_{j}}
\enan


\subsection{The Infinite HMM and Sticky HDP-HMM}

Extensions to the HDP-HMM which explicitly encourage state persistence are the {\em infinite HMM} (iHMM) \cite{Beal:2002} and the {\em sticky HDP-HMM} \cite{Fox:2010tg}. In these models, an extra state self-transition bias parameter $\kappa$ is introduced. Larger values of $\kappa$ bias the state self-transition probability to be close to one: $\pi_{ii} \approx 1$.  In finite Bayesian HMMs this is done by adding $\kappa$ to the $i$-th parameter of the Dirichlet prior for the $i$-th conditional distribution
$\bsym\pi_{i} \sim \Dir(\alpha_{1}\beta_{1},\dots,\alpha_{1}\beta_{i}+\kappa,\dots, \alpha_{1}\beta_{M}).$ Intuitively the infinite HMM case is similar.

To do this in the infinite case the sticky HDP-HMM extends the stick-breaking construction of the HDP-HMM while the iHMM augments the Chinese restaurant franchise representation. 
The only difference between the HDP-HMM and the sticky HDP-HMM generative models is a modification to how the transition distributions $\pi_{i}$ are generated. Specifically, \eqref{eq:hdp-hmm-pi} is replaced by
$
\bsym\pi_{i} | \bsym\beta, \alpha_{1}, \kappa \sim \textnormal{DP}(\alpha_{1}\bsym\beta + \kappa\delta_{i}).
$
The data generation and state transition mechanisms are otherwise the same as the HDP-HMM. The hyperparameter $\kappa$ can be thought of as an extra pseudo-count term on the self-transitions. 

Note that $\kappa$ does not directly parameterize state dwell time.  Call the expected value of $\pi_{ii}$ in the infinite state limit $\rho$, i.e.~$\mathbb{E}[\pi_{ii}] = \frac{\kappa}{\alpha_1 + \kappa} = \rho$.  The expected dwell duration of state $i$ is then $1/({1-\rho})$.  As the expected dwell duration of state $i$ involves $\alpha_1$ and $\kappa$, Fox et al. \cite{Fox:2010tg} propose a reparameterization of the model in terms of $\rho$ and $\alpha_1 + \kappa.$  Note that the global $\kappa$ could in theory be replaced with state-specific parameters $\kappa_{i}$, although it is not straightforward to do so.  This would allow greater heterogeneity in the dwell-time distribution of inferred states.


\subsection{The HDP Hidden Semi-Markov Model}

Another approach to state persistence is the {\em HDP hidden semi-Markov model} (HDP-HSMM)\cite{Johnson:2013}\cite{Johnson:2010}. The HDP-HSMM is a modified version of the HDP-HMM in which states are imbued with state-specific duration distributions.  This, like in the parametric HSMM, requires that self-transitions are precluded. 
To do this the HDP-HSMM supplements \eqref{eq:hdp-hmm-gem}-\eqref{eq:hdp-hmm-theta} with
\eqn \tilde{\pi}_{ij} | \bsym\pi_{i} = \frac{\pi_{ij}\mathbb I(i \ne j)}{1 - \pi_{ii}}. \enn
Just as in the HSMM, $\tilde{\bsym\pi}_{i}$ is the renormalized version of $\bsym\pi_{i}$ after forcing the self-transition probability $\pi_{ii}$ to be zero. The data generation and state transition mechanism of the HDP-HSMM is the same as for the HSMM, as defined in \eqref{eq:state-trans-dynamics}-\eqref{eq:duration-trans-dynamics}. 

The HDP-HSMM offers more flexibility than the sticky HDP-HMM since the state dwell duration distribution remains geometrically distributed in the sticky HDP-HMM, while the HDP-HSMM can make use of arbitrary state dwell duration distributions, depending on what is required for the task. For example, in some applications a Poisson, negative binomial, or even an offset geometric duration distribution might be more appropriate choice for modeling the phenomena of interest. 

\section{The Infinite Structured Hidden Semi-Markov Model}
\label{SIHMM}

The infinite structured hidden semi-Markov model (\SIHMM) is a novel Bayesian nonparametric framework for generating infinite HMMs with explicit state duration distributions and structured transition constraints.   Conceptually it is very closely related to the sticky HDP-HMM and the HDP-HSMM. However, relative to them we claim it has some advantages.  Like the HDP-HSMM, the \SIHMM directly parameterizes state duration distributions allowing for more heterogeneity and specificity in state dwell durations that the sticky HDP-HMM.  This issue can be addressed partially by adding additional state-specific sticky parameters to the sticky HDP-HMM.   Relative to the HDP-HSMM the \SIHMM framework gives rise to different inference algorithms.  Establishing the relative merit of these algorithms is one line of future work.  Most significantly, however, the \SIHMM framework allows one to build models like the infinite left-to-right explicit-duration HMM (ILRHMM).  It is unclear how one would derive models like the ILRHMM from either the HDP-HSMM or the sticky HDP-HMM.


The \SIHMM is the same as the HDP-HSMM except in the way it constructs dependent, infinite-dimensional transition distributions with structural zeros.  The \SIHMM uses spatial normalized gamma processes (SN$\Gamma$Ps)\cite{Rao:2009} to do this. This construction is capable of generating infinite dimensional transition distributions that are structured in the sense that subsets of states can flexibly be made unreachable from others.

\begin{figure}[t]
%
\centering
\includegraphics[width=.5\textwidth]{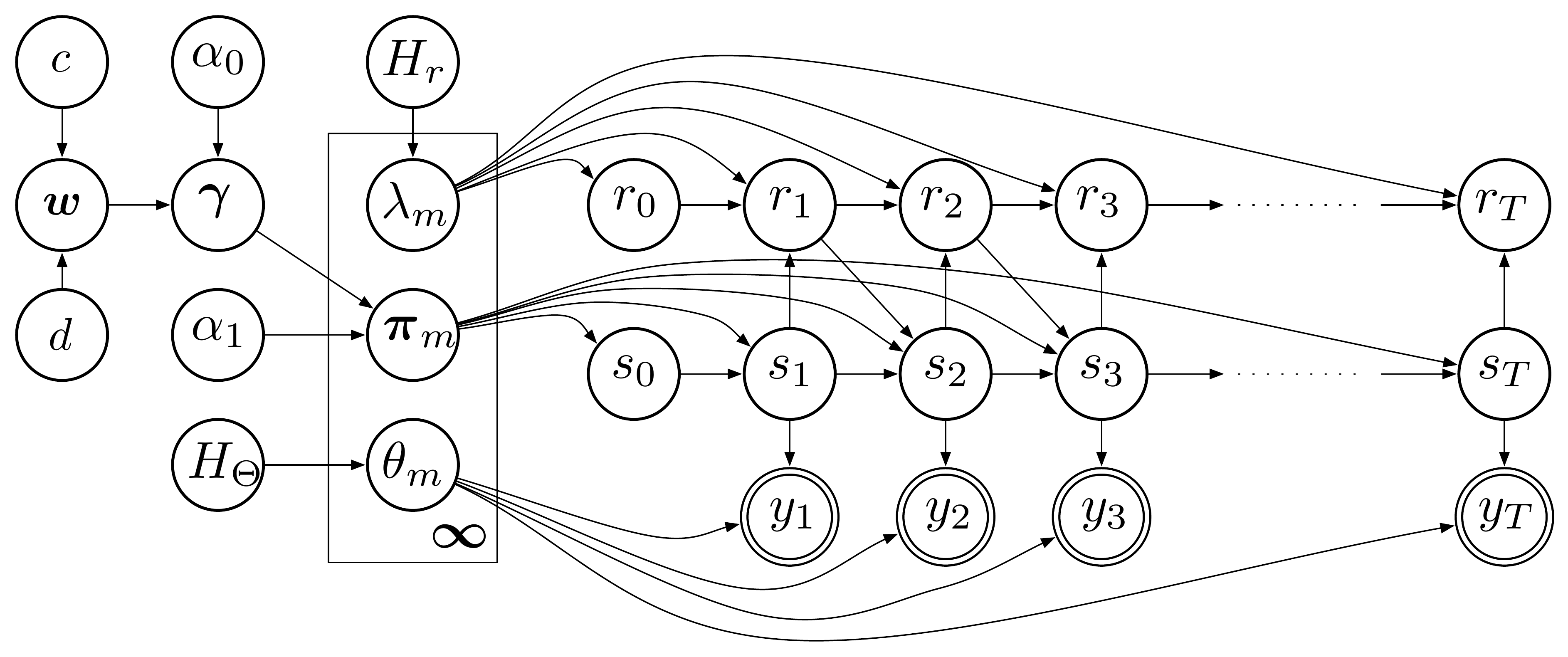}
\caption{Graphical model for the \SIHMM when a Pitman-Yor process is used to define $\tilde H_{\Theta}$ (auxiliary variables for slice-sampling not shown).  }
\label{fig:graphical-model}
\end{figure}
%

%
%
%
\subsection{Infinite Structured Transition Distributions via Spatial Normalized Gamma Processes}
\label{sec:inf_trans_distributions}

The SN$\Gamma$P relies on the gamma process (\GP) construction of the DP \cite{Kingman:1993}. Let $(\Theta, \mathcal B)$ be a measure space and $\mu$ be a totally finite measure defined on that space. A {\em gamma process with base measure $\mu$}, $\Gamma\mathrm{P}(\mu)$, is a random measure on $(\Theta, \mathcal B)$. A draw $G \sim \mathGP(\mu)$ is a measure of the form $G = \sum_{i = 1}^{\infty} \gamma_{i}\delta_{\theta_{i}}$ such for all $\tilde \Theta \in \mathcal B$, 
\eqn G(\tilde\Theta) \sim \GammaRV(\mu(\tilde\Theta), 1). \enn
In general $G$ is {\em not} a probability measure. However, a draw from a DP, which is a probability measure, can be constructed from a \GP draw by normalizing $G$:
\eqn D = G/G(\Theta) \sim \DP(\alpha_{0}H_{\Theta}), \enn
where $\alpha_{0} = \mu(\Theta)$ and $H_{\Theta} = \mu/\alpha_{0}$. This is possible since if $\mu$ is totally finite than $G$ is almost surely totally finite. 

$\mathGP$s have an infinite divisibility property similar to that of gamma random variables: the sum of independent $\Gamma$Ps is also a $\Gamma$P \cite{Kingman:1993}. For example, let $(\Theta_{1}, \Theta_{2},\dots,\Theta_{n})$ be a finite measurable partition of $\Theta$ and let $\mu_{\Theta_{i}}$ be the restriction of $\mu$ to $\Theta_{i}$: $\mu_{\Theta_{i}}(\tilde\Theta) = \mu(\tilde\Theta \cap \Theta_{i})$ for all $\tilde\Theta \in \mathcal B$. Then if $G_{i} \sim \mathGP(\mu_{\Theta_{i}})$, $G := \sum_{i=1}^{n} G_{i} \sim \mathGP(\mu)$. Thus, we can construct many different, related $\Gamma$Ps by summing together subsets of the $\Gamma$Ps defined on $\Theta_{1},\dots,\Theta_{n}$.


The first step in defining an \SIHMM is to choose its state transition topology.  One does so by choosing the set of states $V_m \subset \mathbb N$ to which state $m$ can transition (for $m \in \mathbb N$, the natural numbers). In other words, if $k \notin V_{m}$, then transitions from state $m$ to state $k$ are not permitted: $\pi_{mk} = 0$. For example, in the case of the HDP-HSMM, the ``restricting sets'' would be defined as $V_m = \mathbb N \setminus \{ m \}$ to ensure $\bsym \pi$ has the structural zeros $\pi_{mm} = 0$. Figure \ref{fig:graphical-model} shows the \SIHMM graphical model.

Spatial normalized gamma processes (SN$\Gamma$Ps)  \cite{Rao:2009} can be used to construct the infinite-dimensional conditional transition distributions in such a way that they have the desired structural zeros as defined by $V_{m}$.
An overview of how SN$\Gamma$Ps can be used to generate structurally constrained transition distributions for the \SIHMM is as follows
\begin{list}{\labelitemi}{\leftmargin=1em}
\item Draw an unnormalized marginal state occupancy distribution from a base $\Gamma$P defined over the entire set of states.   Restrict this distribution according to $V_m$ for each state.  Normalize to produce a DP draw $D_m$ for each state.  This is like the base transition distribution in the HDP-HMM but restricted to the states accessible from $m$.
\item Generate the transition distribution $\bsym \pi_m$ from $D_m$.  
\end{list}

We start the formal treatment of this procedure by defining a base measure for the base \GP of the form $\mu = \alpha_{0}H_{\Theta\times\mathbb N}$, where $\alpha_{0}$ is a concentration parameter and $H_{\Theta \times \mathbb N}$ is a probability measure on the joint space of parameters $\Theta$ and states $\mathbb N$. To construct $H_{\Theta \times \mathbb N}$, we must construct an atomic random measure $\tilde H_{\Theta}$ over the parameter space $\Theta$. A straightforward way to accomplish this is to let $\tilde H_{\Theta}$ be the realization of a Pitman-Yor process with mean measure $H_{\Theta}$, concentration parameter $c$, and discount parameter $d$: $\tilde H_{\Theta} \sim \mathcal{PY}(c,d,H_{\Theta})$. This is the approach taken in our experiments, though it is certainly not the only possible one. In any case, without loss of generality we write  
 $\tilde H_{\Theta} = \sum_{m=1}^{\infty} w_{m} \delta_{\theta_{m}}$. If the Pitman-Yor process approach is used, then $\bsym w \sim \mathcal{SB}(c,d)$ is drawn from the two-parameter stick-breaking prior \cite{Ishwaran:2001} and $\theta_m \sim H_{\Theta}$.  Let 
\eqan
\mu(\tilde \theta, \tilde M) &=& \alpha_{0} H_{\Theta\times\mathbb N}(\tilde \theta, \tilde M) \\
&=& \alpha_{0} \sum_{m=1}^{\infty} w_{m} \mathbb{I}({\theta_{m}} \in \tilde \theta) \mathbb I({m} \in \tilde M).
\enan
This associates a single observation distribution parameter $\theta_m$ and weight $w_m$ with each state $m$.  The introduction of the additional measure on the product space $\Theta \times \mathbb N$ may at first appear unnecessarily complex. However, as explained in more detail below, it is critical in order to use the SN$\Gamma$P construction. 

To generate transition distributions for each state that conform to our choice of $V_m$ requires some organizing notation.  Partition $\mathbb N$ into a collection of {\em disjoint} sets $\mathcal A$ such that every $V_m$ can be constructed by taking the union of sets in $\mathcal A$.  Let $\mathcal A_m$ be the collection of sets in $\mathcal A$ whose union is $V_m$. Let $R$ be an arbitrary element of $\mathcal A.$  This partitioning is necessary to keep track of the independent DPs that will be mixed below in \eqref{eq:DPmixture}.   These conditions do not uniquely define $\mathcal A$. So while any partition that satisfies these conditions will work, some choices can result in more computationally efficient inference.

Consider a gamma process $G \sim \Gamma\textnormal{P}(\mu)$ with base measure $\mu$. For any subset $S \subset \mathbb N$, define the ``restricted projection'' of $\mu$ onto $S$ as $\mu_{S}(\tilde\Theta) = \alpha_{0}H_{\Theta \times \mathbb N}(\tilde\Theta, S)$. Note that $\mu_{S}$ is a measure over the space $\Theta$ instead of over $\Theta \times \mathbb N$. Define the restricted projection of $G$ onto $S$ to be $G_{S}(\tilde\Theta) = G(\tilde\Theta, S) \sim \mathGP(\mu_{S})$ so that $G_{S}$ is \GP distributed with base measure $\mu_{S}$. This follows from the fact that restrictions and projections of {\GP}s are {\GP}s as well. Let $D_{S} = G_{S}/G_{S}(\Theta) \sim \DP(\mu_{S})$ be the DP that arises from normalizing $G_{S}$. 

We are interested in constructing the set of dependent DP draws $D_{m} := D_{V_{m}}$ from the restrictions of $G$ to the sets $V_{m}$. Clearly, for disjoint sets $S, S' \subset \mathbb N$, $G_{S \cup S'} = G_{S} + G_{S'}$, so $G_{V_{m}} = \sum_{R \in \mathcal A_{m}} G_{R}$. Therefore, 
\eqan
D_{{m}} &=& D_{V_{m}} = \frac{G_{V_{m}}}{G_{V_{m}}(\Theta)} = \frac{\sum_{R \in \mathcal A_{m}} G_{R}}{\sum_{R \in \mathcal A_{m}} G_{R}(\Theta)} \nonumber \\
&=& \sum_{R \in \mathcal A_{m}} \frac{G_{R}(\Theta)}{\sum_{R' \in \mathcal A_{m}} G_{R'}(\Theta)} D_{R} \nonumber \\
&=& \sum_{R \in \mathcal A_{m}} \frac{\gamma_R}{\sum_{R' \in \mathcal A_{m}} \gamma_{R'}}  D_{R}\label{eq:DPmixture}
\enan
Here, by the definition of the \GP
\eqn \gamma_{R} \sim \GammaRV(\mu_{R}(\Theta), 1). \label{eq:gamma-sngp}\enn
Thus, $D_{m}$ is actually a mixture of independent DPs. 

The $D_{m}$ are tightly coupled draws from dependent Dirchlet processes.  Intuitively these play the same role as $\bsym \beta$ in the HDP-HMM but with one important difference.  Here $D_{m}$ has zero weight on all atoms corresponding to unreachable states. These $D_{m}$ then serve as base distributions for drawing the conditional state transition distributions $\tilde D_{m}\sim\textnormal{DP}(\alpha_{1}D_{m})$, where $\tilde D_{m} := \sum_{k=1}^{\infty}\pi_{mk}\delta_{\theta_{k}}$. The resulting vectors $\bsym\pi_{m} = (\pi_{m1},\pi_{m2},\pi_{m3},\dots)$ are infinite state transition distributions with structural zeros.  Drawing them in this way allows each state to flexibly deviate from the restricted, normalized marginal state occupancy distribution $D_{m}$ yet still share statistical strength in a hierarchical manner. 

By choosing particular sets $V_m$ {\SIHMM}s encode different kinds of transition structures into infinite HMMs.  In the following two sections we describe choices that lead to infinite variants of two common finite HMM types.  The most important difference between the two models is the way in which the set of states is partitioned.

The necessity of constructing $H_{\Theta\times\mathbb N}$, and in particular doing so from the atomic random measure $\tilde H_{\Theta}$, can now be understood. The $\mathbb N$ portion of the product space $\Theta\times\mathbb N$ acts as the auxiliary space for the SN$\Gamma$P construction. By only considering a subset of  $\mathbb N$ (namely $V_{k}$) we are able to eliminate some of the atoms $\Theta$ space in a controlled manner. That is, consider $S = \mathbb N$ and the restricted projection $\mu_{S}$. Each time some element $k \in S$ is removed from $S$, $\mu_{S}$ will have no longer have the atom $\delta_{\theta_{k}}$, thus eliminating the $k$-th state from the transition structure. But, note that we want the parameter $\theta_{k}$ of the $k$-th state to be random, hence the requirement that $H_{\Theta \times \mathbb N}$ be a random measure.

\subsection{The Infinite Explicit Duration HMM}
\label{sec:IED-HMM}
The {\em infinite explicit duration HMM} (IED-HMM) disallows self-transitions in order to support explicitly parameterized state dwell duration distributions. The IED-HMM yields a nearly equivalent model to the HDP-HSMM. However, the \SIHMM construction gives rise to different inference algorithms.  We maintain separate naming conventions as in our experiments the general purpose \SIHMM inference algorithms are used. 

Let the range of each state $m$ (the states reachable from $m$) be $V_{m} = \mathbb N \setminus \{ m \}$. With this choice of range we can complete the generation of infinite transition distributions for an IED-HMM.  To simplify the discussion of inference in this model we arbitrarily partition the infinite set of states into $\mathcal T \subset \mathbb N$ and  $R_{+} = \mathbb N \setminus \mathcal T$. 

Given the choice of $V_{m} = \mathbb N \setminus \{ m \}$, we have 
\eqn \mathcal A = \{R_{+}\} \cup \{ \{ m \} | m \in \mathcal T \}, \qquad \mathcal A_{m} = \mathcal A \setminus \{ m\}. \enn
To simply notation, let $\gamma_{m} := \gamma_{\{m\}}$, $\gamma_{+} := \gamma_{R_{+}}$ and $w_{+} := \sum_{k' \in R_+}w_{k'}$. With $m \in \mathcal T, k \in \mathbb N$, plugging into \eqref{eq:gamma-sngp} gives 
\eqan
\gamma_{m} &\sim&  \mathsf{Gamma}(\alpha_{0}w_{m}, 1) \label{eq:ied-gamma-m} \\
\gamma_{+} &\sim&  \mathsf{Gamma}(\alpha_{0}w_{+}, 1)   \label{eq:ied-gamma-plus}
\enan
since $\mu_{\{m\}}(\Theta) = \alpha_{0}w_{m}$ and $\mu_{R_{+}}(\Theta) = \alpha_{0}w_{+}$. Since $D_{\{k'\}} = \delta_{\theta_{k'}}$, plugging into  \eqref{eq:DPmixture} gives
\eqan
D_{{m}} &=&  \sum_{R \in \mathcal A_{m}} \frac{\gamma_R}{\sum_{R' \in \mathcal A_{m}} \gamma_{R'}}  D_{R} \\
&=& \sum_{k' \in \mathcal T  \cap V_m} \beta_{mk'} \delta_{\theta_{k'}} + \beta_{m+}  D_{{+}}
\enan
where
\eqan
\beta_{mk}  &=&  \frac{\mathbb I(k \in \mathcal T \cap V_m)\gamma_{k}}{\gamma_{+} + \sum_{k' \in \mathcal T \cap V_m} \gamma_{k'}} \\
\beta_{m+}  &=& \frac{\gamma_{+}}{\gamma_{+} + \sum_{k' \in \mathcal T \cap V_m} \gamma_{k'}} \label{eq:beta} \\
D_{{+}} &\sim& \textnormal{DP}(\mu_{R_{+}}) \label{eq:D_m}.
\enan

If $\beta_{mk} \neq 0$ then state $k$ is reachable from state $m$. 
Denoting $\bsym\beta_{m} = (\beta_{m1}, \dots, \beta_{mM}, \beta_{m+})$, draw the structured state transition distributions $\bsym\pi_{m}= (\pi_{m1}, \dots, \pi_{mM}, \pi_{m+})$ from 
\eqn \bsym\pi_{m} \sim \mathsf{Dirichlet}(\alpha_{1}\bsym\beta_{m}). \label{eq:pi} \enn
The conditional state transition probability row vector $\bsym\pi_{m}$ is finite dimensional only because the probability of transitioning to the states in the set $R_+$ is explicitly summed: $\pi_{m+} = \sum_{k'  \in R_+} \pi_{mk'}$.  During inference it is often necessary to perform inference about states that are part of $R_+$. Sec.~\ref{sec:forward_consideration} explains how to dynamically grow and shrink $R_+$ as needed.

\subsection{The Infinite Left-to-Right HMM}

Our method of generating structured transition distributions allows one to easily specify other kinds of structured infinite HSMMs.  An infinite left-to-right HSMM (ILR-HMM) requires that transitions to previously used states be forbidden, namely~$\pi_{mk} = 0$ for all $k \le m$. To the best of our knowledge, no nonparametric generalization of the left-to-right HMM has been defined before now.  Enforcing the left-to-right condition on $\pi_{mk}$ in the \SIHMM simply requires defining the restricting sets to be of the form $V_{m} = \{ m+1,m+2,m+3,\dots \}$. This choice of $V_{m}$ requires increased notational complexity but does not lead to changes to the generative model. 

For state $m \in \mathcal T$, let $m^{*}$ be the smallest $m' \in \mathcal T$ such that $m' > m$, or $\infty$ if no such $m'$ exists. Let $R_{m+} = \{ k' \in \mathbb N | m < k' < m^{*} \}$ be the region in between $m$ and the next index that is in $\mathcal T$ (i.e.~$m^{*}$). Then 
\eqan
\mathcal A &=& \{ \{ m \} | m \in \mathcal T \} \cup \{ R_{m+} | m \in \mathcal T \} \\
\mathcal A_{m} &=& \{ \{ m' \} | m' \in \mathcal T_{m'}, m' > m \} \cup \\
&& \{ R_{m'+} | m' \in \mathcal T, m' \ge m \}. \nonumber
\enan
Hence, letting $\gamma_{m+} := \gamma_{R_{m+}}$ and for $m \in \mathcal T$ and $k \in \mathbb N$, plugging into \eqref{eq:gamma-sngp} gives
\eqan
\gamma_{m} &\sim&  \mathsf{Gamma}(\alpha_{0}w_{m}, 1) \label{eq:ilr-gamma-m}\\
\gamma_{m+} &\sim& \mathsf{Gamma}(\alpha_{0}\textstyle\sum_{k \in R_{m+}}w_{k}, 1) \label{eq:ilr-gamma-m-plus}
\enan
while plugging into \eqref{eq:DPmixture} gives
\eqan
D_{{m}} &=& \beta_{mm+}  D_{{m+}}  + \nonumber \\
&& \textstyle\sum_{k \in \mathcal T \cap V_m} (\beta_{mk} \delta_{\theta_{k}} + \beta_{mk+}  D_{{k+}}) 
\enan
where
\eqan
\beta_{mk} &=& \frac{\mathbb I(k \in \mathcal T \cap V_m)\gamma_{k}}{\gamma_{m+} + \sum_{k' \in \mathcal T \cap V_{m}}(\gamma_{k'} + \gamma_{k'+}) } \\
\beta_{mk+} &=& \frac{\mathbb I(k \in (\mathcal T \cap V_m) \cup \{ m \})\gamma_{k+}}{\gamma_{m+} + \sum_{k' \in \mathcal T \cap V_{m}}(\gamma_{k'} + \gamma_{k'+})}  \label{eq:beta-ilr}  \\
D_{{m+}} &\sim& \textnormal{DP}(\mu_{R_{m+}}). \label{eq:D-m-ilr}
\enan
The $\bsym \pi_m$'s are drawn as they are in the IED-HMM. 

%
\othermodels{
\subsection{Relation to Other Models}
\label{sec:model-comparison}
%

{\bf HDP-HSMM.} The difference between the IED-HMM and the HDP-HSMM is that the IED-HMM has an extra level in the graphical model: $\bsym w, c$, and $d$ are not present in the HDP-HSMM. Using the Pitman-Yor process (PYP) construction of the atomic measure $\tilde H_{\Theta}$ the HDP-HSMM is recovered by (a) setting the discount parameter of the PYP to be $d = 0$ so $\bsym w \sim \GEM(c)$ and (b) letting the concentration parameter $\alpha_{0} \to \infty$ so that the ``variance'' about $\bsym w$ is zero, forcing $\bsym \gamma = \bsym w$ with probability 1.   With this simplification, the IED-HMM and the HDP-HSMM result in mathematically equivalent models.  

{\bf HDP-HMM.} The HDP-HMM can be recovered by choosing the duration distribution to be the delta function at zero, $H_r = \delta_0$, so that the duration counter is always equal to zero, $r_t = 0$, for all $t$. This restores the implicit geometric state duration distribution of the HDP-HMM. 

Also, because all possible state transitions are permitted in the HDP-HMM, the restricting sets $V_{m}$ on the auxiliary space should be chosen such that the dependent {\GP}s are all equal, i.e.~by setting $V_m = \mathbb N$. If all the dependent {\GP}s are equal, draws from the dependent DPs are equal as well: $D_{m} = D_{m'}~\forall m, m' \in \mathcal T$. In this case, all of the conditional state transition distributions $\tilde D_{m}$ are independent draws from DPs with the same base measure, exactly like the HDP-HMM. 

{\bf Finite Bayesian HMM.} To recover the standard, finite Bayesian HMM with $K$ states, use the same parameter setup as for the HDP-HMM, but replace the usual auxiliary space $\mathbb N$ everywhere by a finite one, $\mathbb S := [K]$. Setting $V_{m} = \mathbb S$, all of the DP draws become Dirichlet distributed instead since they will only have $K$ atoms. 

{\bf Sticky HDP-HMM.} While the sticky HDP-HMM cannot be exactly reproduced in the \SIHMM framework, a very similar model can be created using the IED-HMM (or, equivalently, the HDP-HSMM) by using a geometric distribution 
for the duration distribution $F_{r}$. %

{\bf Other HMM and change-point models.} Learning of HMMs with an unknown number of states using reversible jump MCMC has been considered \cite{Robert00}.  This approach allows MCMC sampling of finite HMMs with unknown but finite state cardinality using state split, combine, birth, and death transition transition operators.  Incorporating explicit state duration and transition constraints into such a model might be possible but would require designing complex analogues to these operators.  

An auxiliary variable sampling scheme developed for learning finite HMMs where each state's emission distribution is an infinite mixture \cite{Yau11} is methodologically related to the slice sampling approach we employ. In particular, in \cite{Yau11} the emission mixture component responsible for generating an observation is selected via an auxiliary variable scheme that restricts this choice to a finite subset of the infinite set of possible mixture components.

A multivariate time series segmentation model based on the product partition model (PPM; \cite{Barry93}) \cite{Xuan07} is closely related to our ILR-HMM.  The PPM is essentially an HMM where the latent quantities of interest are the times at which changepoints occur rather than the state identity at each time.  The states in our ILR-HMM are equivalent.  The PPM as described has a single geometric distribution on segment length, though this restriction could presumably be lifted to arrive at a model similar to our ILR-HMM.  Inference in the two models is different: PPM inference is performed using a $O(T^2)$ dynamic programming approach \cite{Fearnhead07} which exactly computes the map partition (and implicitly the partition cardinality K and per-segment observation parameter distributions).  There is an approximate inference algorithm which scales like our $O(T)$ approach \cite{Fearnhead07}, but it does not have the kinds of asymptotic convergence guarantees the auxiliary variable slice sampling approach does \cite{VanGael:2008}.

}
\section{\SIHMM Inference}
\label{sec:inference}

{\SIHMM} inference requires sampling the hidden state and duration variables, $\bsym s$ and $\bsym r$, the transition matrix, $\bsym \pi$, and the higher level weights $\bsym \beta$ and $\bsym w$. In addition, the emission and duration distribution parameters, $\bsym \theta$ and $\bsym \lambda$, as well as the hyperparameters must be sampled. We take a blocked Gibbs sampling approach similar to standard approaches to inference in the HDP-HMM and HDP-HSMM. In particular, to jointly sample the states $\bsym s$ and durations  $\bsym r$ conditioned on the others we employ the forward filtering backward slice sampling approach of \cite{VanGael:2008} for infinite HMMs and \cite{Dewar:2012} for explicit duration HMMs. 
\begin{figure*}[tbp]
\centering
\begin{subfigure}[b]{.33\textwidth}
\centering
\includegraphics[width=\textwidth]{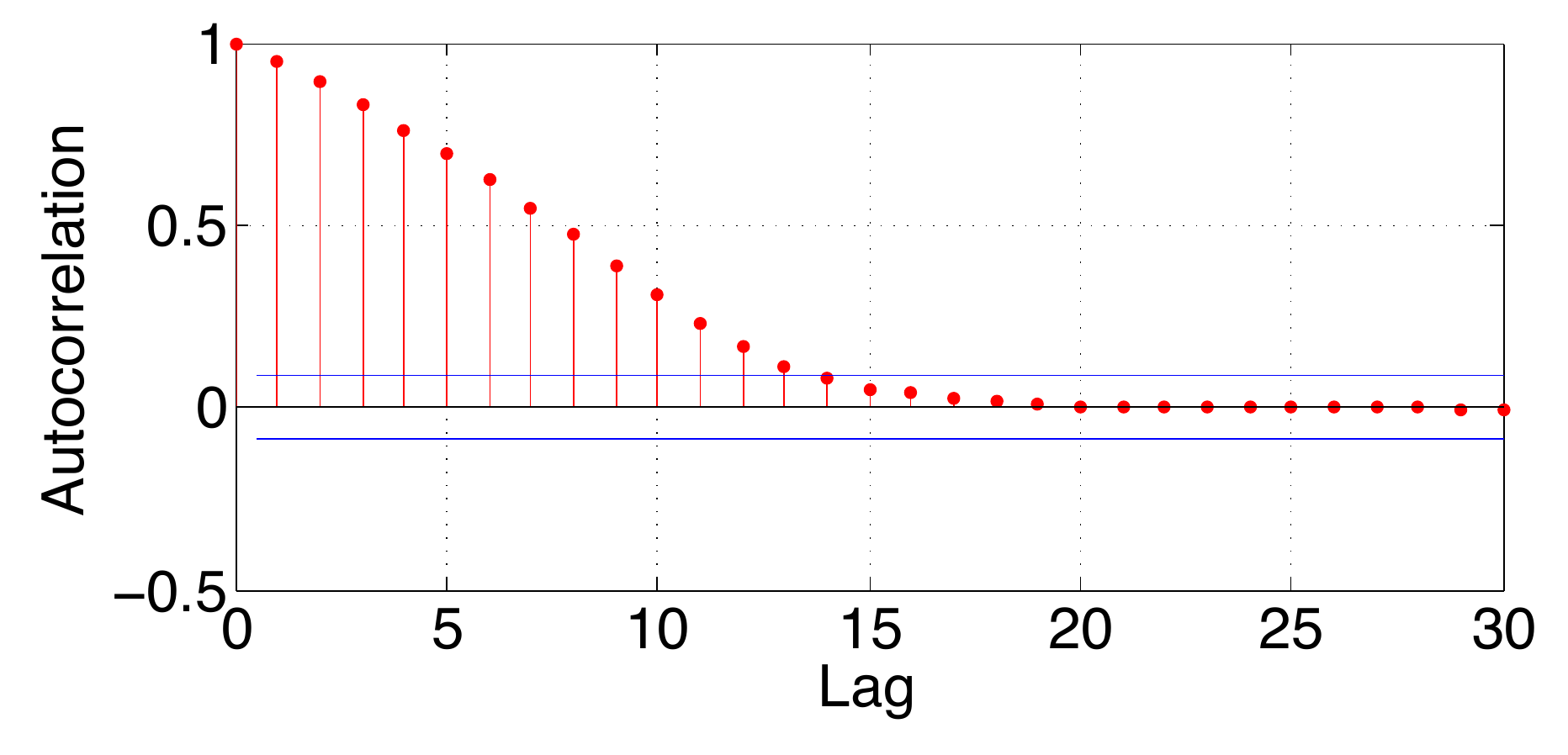}
\caption{}
\end{subfigure}
\begin{subfigure}[b]{.33\textwidth}
\centering
\includegraphics[width=\textwidth]{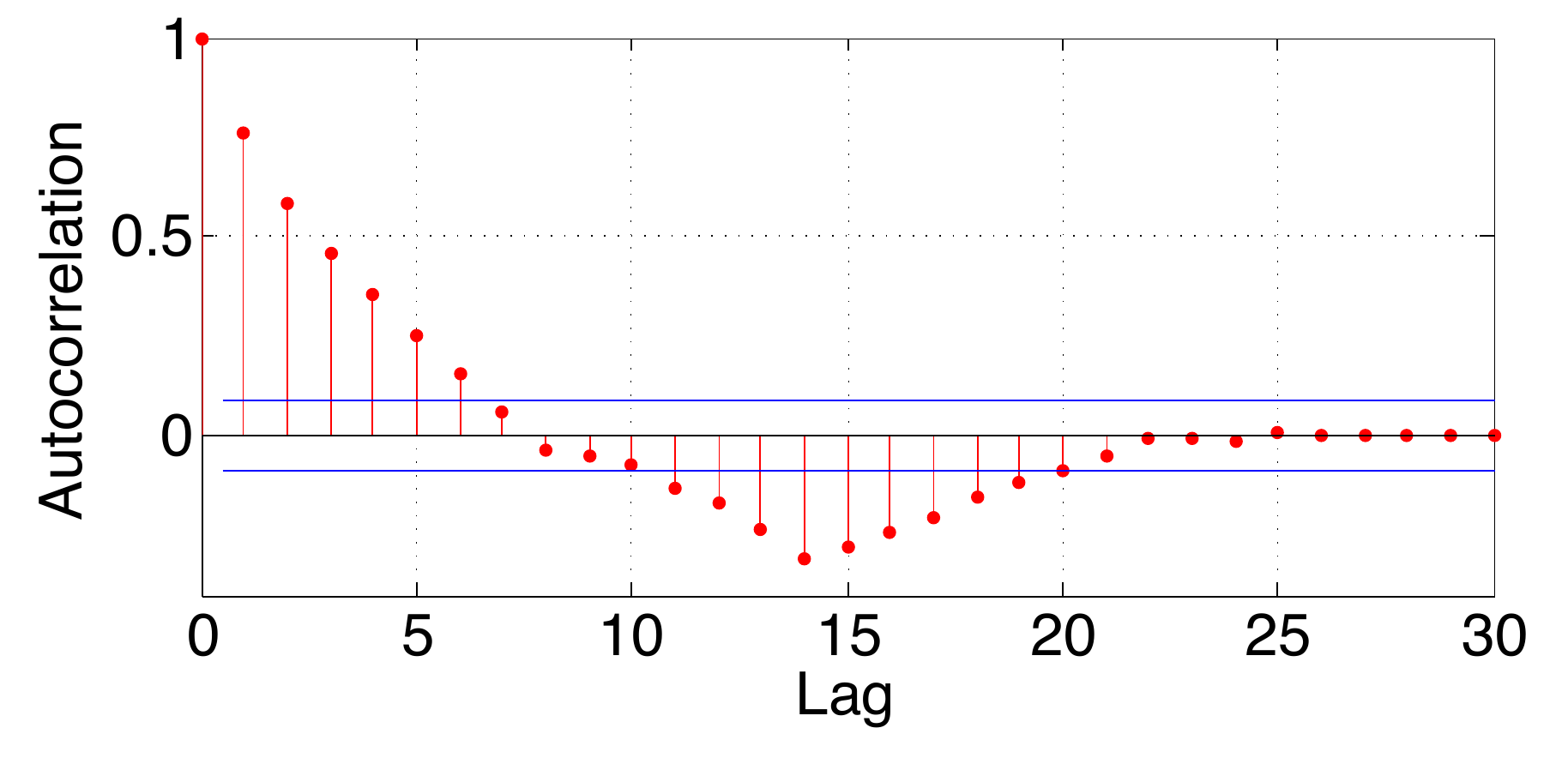}
\caption{}
\end{subfigure}
\begin{subfigure}[b]{.32\textwidth}
\centering
\includegraphics[width=\textwidth]{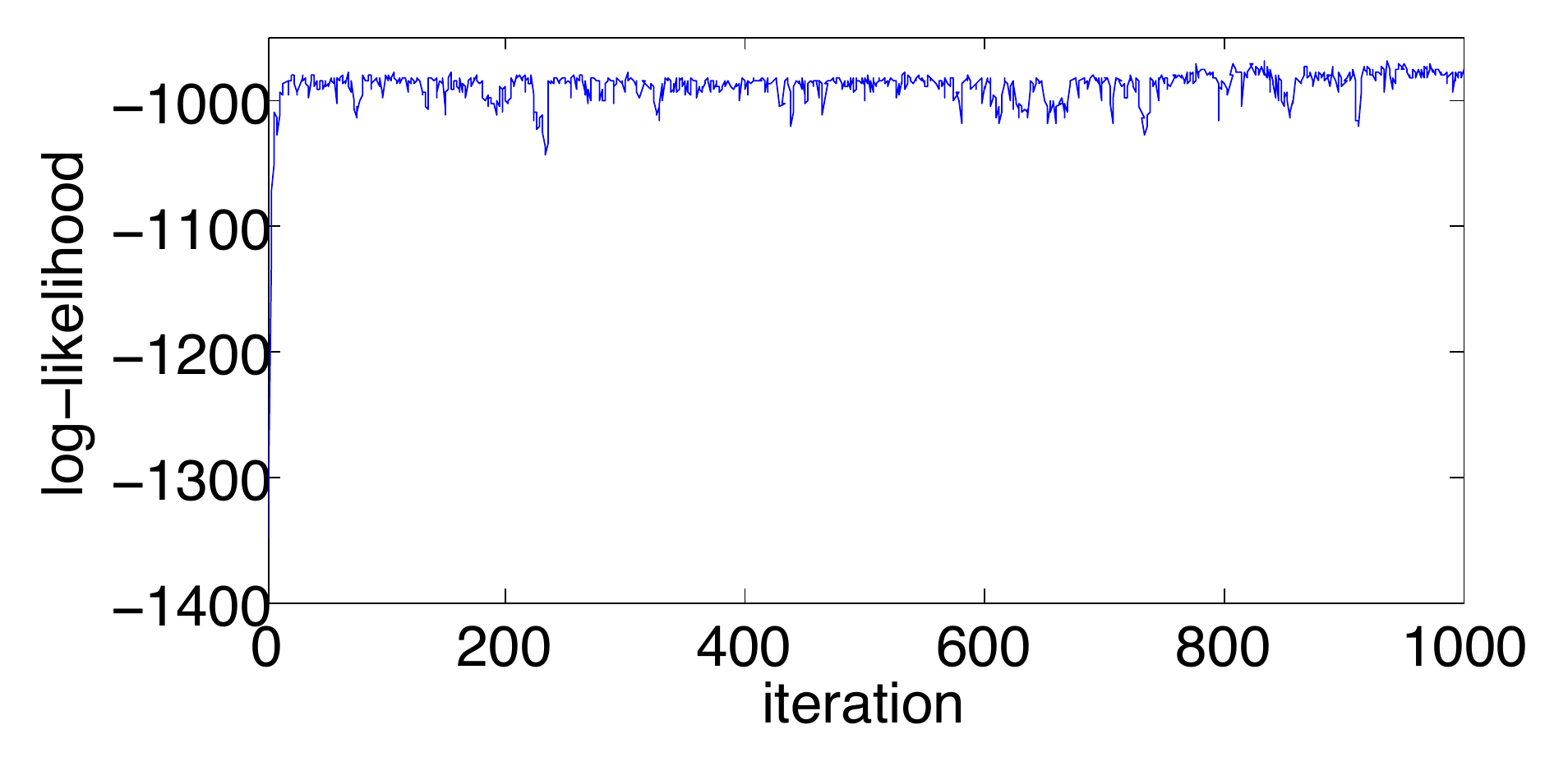}
\caption{}
\end{subfigure}
\caption{{\bf (a)} and {\bf (b)} Worst (most slowly decreasing) IED-HMM sample autocorrelations (data from Fig.~\ref{fig:overlaid-means-and-state-distribution}). The autocorrelations are for the mean and duration parameters, respectively, associated with the sampled state that generated an observation at a single fixed time. {\bf (c)} Joint log-likelihood plot for the first 1000 iterations of the IED-HMM sampler. The sampler converges and mixes quickly due to the forward-backward slice sampler making large moves in the state space.  }
\label{fig:edihmm-autocorr}
\end{figure*}

{\bf Sampling $\bsym s$ and $\bsym r$ } 
Ideally, the whole trajectory of hidden states $\bsym s$ of an infinite HMM would be resampled at once, as is done by the forward-backward algorithm for parametric Bayesian HMMs. While the infinite number of states makes standard forward-backward impossible since the time complexity of forward-backward in polynomial in the number of states, this difficulty can be overcome by using an auxiliary variable technique known as slice or {\em beam sampling}. The trick is to introduce an auxiliary variable $\bf u$ such that running forward-backward conditioned on it results in only a finite number of states needing to be considered \cite{VanGael:2008}. We first review beam sampling in the context of the HDP-HMM, then generalize it to the \SIHMM. The choice of auxiliary variable distribution is key.  For each time $t$ introduce an auxiliary variable 
\eqn u_{t} |\bsym\pi, s_{t},s_{t-1} \sim \Unif(0, \pi_{s_{t-1}s_{t}}). \label{eq:ut-unif} \enn
Sampling of $\bsym s$ is done conditioned on $\bsym u$ and the other variables. The key quantity for running the forward-backward algorithm is the forward variable $\alpha_{t}(s_{t}) := p(s_{t}|y_{1:t},u_{1:t})$, which, with
\eqn p(u_{t} | \bsym\pi, s_{t}, s_{t-1}) = \frac{\Ind(0 < u_{t} < \pi_{s_{t}s_{t-1}})}{\pi_{s_{t}s_{t-1}}}, \enn
can be computed recursively (the ``forward pass'')
\eqan
\alpha_{t}(s_{t}) &\propto& p(s_{t}, u_{t}, y_{t} | y_{1:t-1},u_{1:t-1}) \nonumber \\
&=& \sum_{s_{t-1}} p(y_{t} |s_{t})p(u_{t}|s_{t}, s_{t-1})p(s_{t}|s_{t-1})\alpha_{t}(s_{t-1}) \nonumber \\
&=& p(y_{t} |s_{t}) \sum_{s_{t-1}} \Ind(u_{t} < \pi_{s_{t}s_{t-1}})\alpha_{t}(s_{t-1}) \nonumber \\
&=& p(y_{t} |s_{t}) \sum_{s_{t-1}: u_{t}<\pi_{s_{t}s_{t-1}}} \alpha_{t}(s_{t-1}) 
\enan
where $p(y_{t}| s_{t}) = F_\theta(y_t | \theta_{s_t})$ and the conditioning on variables other than $\bsym y$, $\bsym s$, and $\bsym u$ have been suppressed for clarity. Since only a finite number of transition probabilities $\pi_{ij}$ can be greater than $u_{t}$, the summation is finite. Intuitively, conditioning on $\bsym u$ forces only the transitions with probability greater than $u_{t}$ to be considered at time $t$ because otherwise $u_{t}$, conditional on that $\pi_{ij}$, could not have been drawn from a distribution whose support is $[0, \pi_{ij}]$. Since $u_{t}$ is less than the probability of the previous state transition at time $t$, there will always be at least some valid path through the states. Given these forward messages, states can be sampled backwards in time from
\eqan
p(s_{T} | y_{1:T}, u_{1:T}) &=& \alpha_{T}(s_{T}) \\
p(s_{t} | s_{t+1}, y_{1:T},  u_{1:T}) &\propto& \alpha_{t}(z_{t})p(s_{t+1} | s_{t}, u_{t+1}).
\enan

Extending this approach to the \SIHMM requires defining a ``full'' state random variable $z_{t} = (s_{t}, r_{t})$ and changing the distribution of the auxiliary random variable $u_{t}$ to 
\eqn p(u_{t} | z_{t}, z_{t-1}) = \mathbb I(u_{t} < p_{t}) p_{t}p_{\beta}(u_{t}/ p_{t}; \alpha_{u}, \beta_{u}), \label{eq:u-prob} \enn
where $p_{\beta}(\cdot; \alpha_{u}, \beta_{u})$ is the density for the beta distribution with parameters $\alpha_{u}$ and $\beta_{u}$ and
\eqan
p_t &:=& p(z_{t} | z_{t-1}) =
 p((s_t , r_t ) | (s_{t-1}, r_{t-1})) = \nonumber \\
	&=&\begin{cases} 
	r_{t-1} > 0, & \mathbb{I}(s_t=s_{t-1})\mathbb{I}(r_t = r_{t-1}-1) \\
	r_{t-1} = 0, & \pi_{s_{t-1}s_{t}}F_r(r_t;\lambda_{s_t}).
	\end{cases} \nonumber 
\enan 
In the case of $\alpha_{u} = \beta_{u} = 1$, equation \eqref{eq:ut-unif} is recovered. It is straightforward to sample $\bsym u$ according to equation \eqref{eq:u-prob}. 
%
%
The forward variables $\alpha_{t}(z_{t})$ then become 
\eqan
\alpha_{t}(z_{t}) &:=& p(z_{t} | y_{1:t}, u_{1:t})  \nonumber \\
&\propto&  p(z_{t}, u_{t}, y_{t}  |  y_{1:t-1}, u_{1:t-1}) \nonumber \\
&=&  \sum_{z_{t-1}}  p(y_{t}| z_{t})p(u_{t} | z_{t}, z_{t-1})p_t p(z_{t-1} | y_{1:t}, u_{1:t}) \nonumber \\
&=& p(y_{t} | s_{t}) \sum_{z_{t-1}} \mathbb I(u_{t} < p_{t})p_{\beta}^\star\alpha_{t-1}(z_{t-1}) \nonumber  \\
&=& p(y_{t} | s_{t}) \sum_{z_{t-1} : u_{t} < p_{t}}  p_{\beta}^\star \alpha_{t-1}(z_{t-1}) \label{eq:forward-var}
\enan 
where $p_{\beta}^\star := p_{\beta}(u_{t}/ p_{t}; \alpha_{u}, \beta_{u})$. 
Samples of the full latent states are taken during the backward pass by sampling from
\eqan
p(z_{T} |  y_{1:T}, u_{1:T}) &=& \alpha_{T}(z_{T}) \\
p(z_{t} | z_{t+1}, y_{1:T},  u_{1:T}) &\propto& p(z_{t}, z_{t+1}, y_{1:T}, u_{1:T}) \nonumber \\
&=& p(u_{t+1} | z_{t+1}, z_{t})p(z_{t+1} | z_{t}) \times \nonumber \\
&& p(z_{t} | u_{1:t}, y_{1:t}) \nonumber \\
&=& \mathbb I(u_{t+1} < p_{t+1})p_{\beta}^\star\alpha_{t}(z_{t}).
\enan
\begin{figure*}[tbp]
\center
\begin{subfigure}[b]{.49\textwidth}
\includegraphics[width=\textwidth]{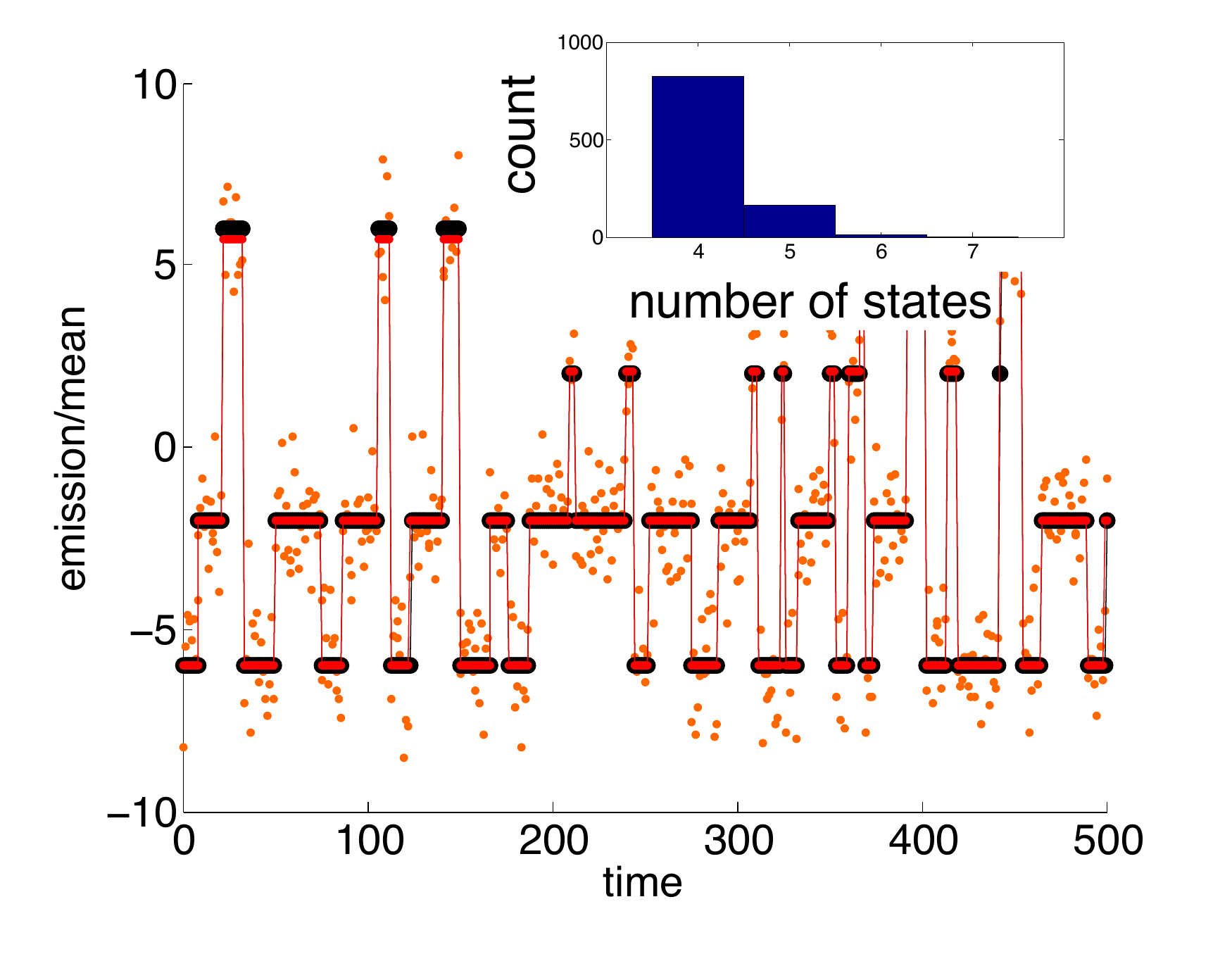}
\caption{}
\label{fig:overlaid-means-and-state-distribution}
\end{subfigure}
~
\begin{subfigure}[b]{.42\textwidth}
\centering
\includegraphics[width=\textwidth]{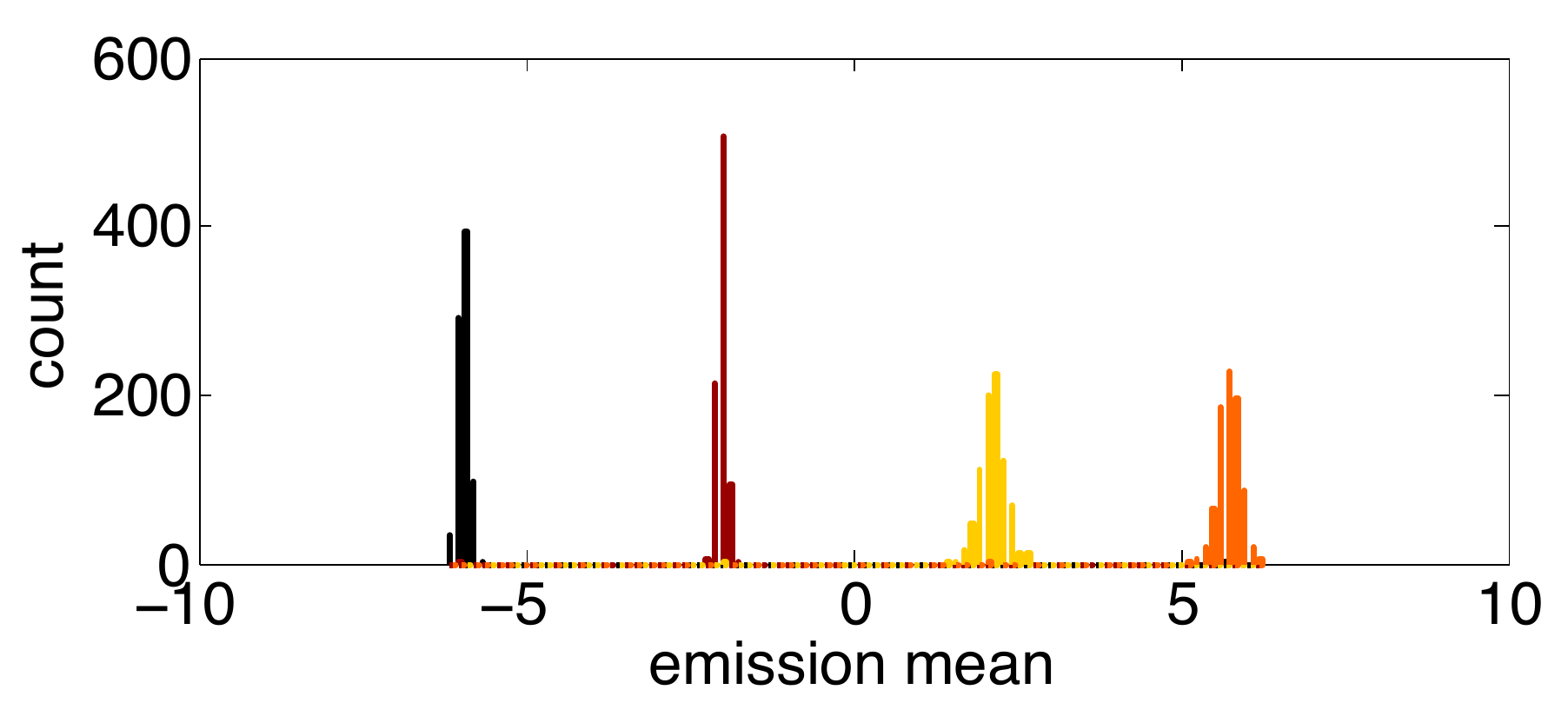}\label{subfig:mean-posteriors-synthetic}  \\
\includegraphics[width=\textwidth]{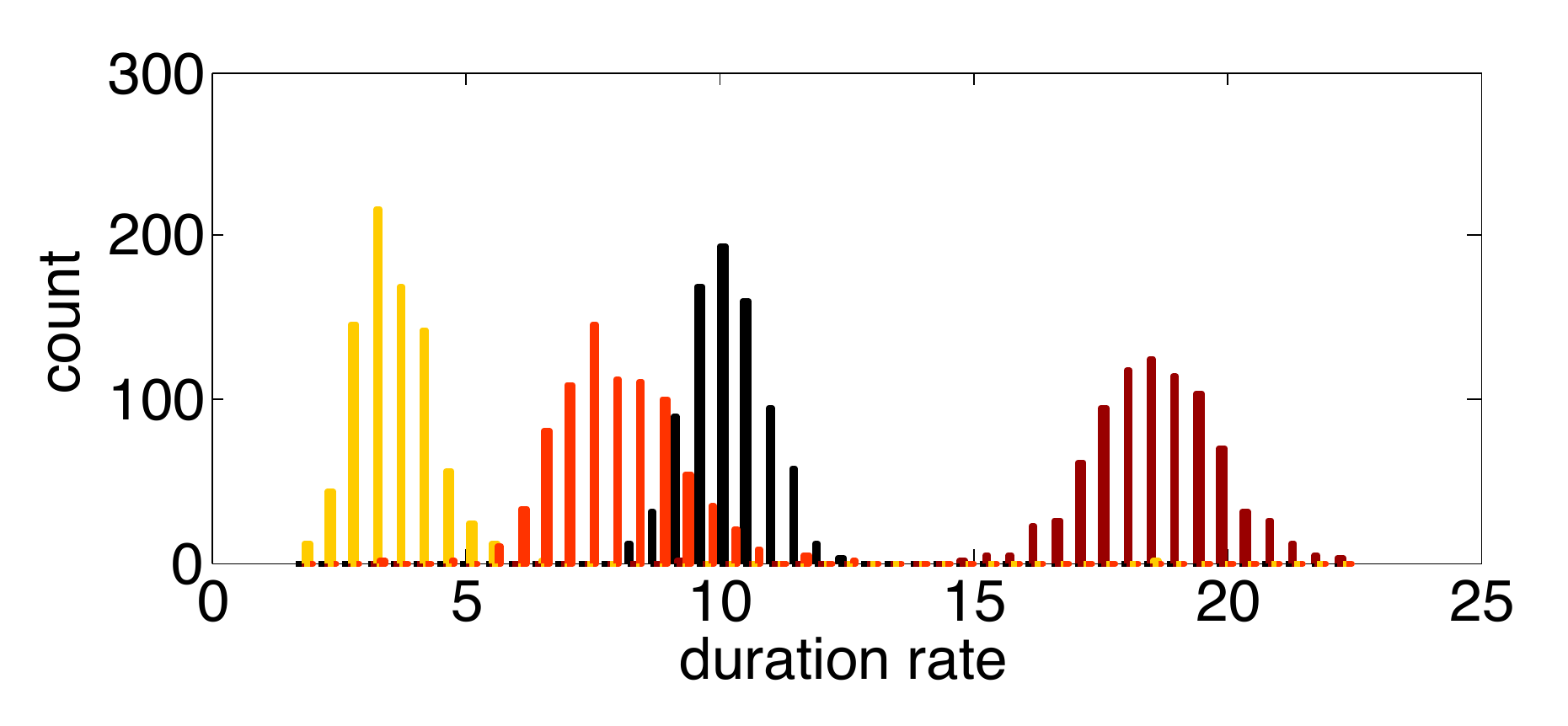}\label{subfig:rate-posteriors-synthetic}
\caption{}
\label{fig:posteriors-synthetic}
\end{subfigure}
\caption{{\bf (a)} Synthetic data generated from a 4 state HSMM with MAP IED-HMM sample overlaid; means of the sampled states (red) are overlaid on the true data (orange) and true means (black). {\em Inset:} Posterior distribution of IED-HMM utilized state counts.  {\bf (b)} Posterior distributions of latent state parameters for data in (a). The true means were -6, -2, 2, and 6 and the true duration rates were 10, 20, 3, and 7. }
\end{figure*}

It is convenient to express $\alpha_{u}$ and $\beta_{u}$ in terms of a single ``temperature'' parameter $\mathcal K$: $\alpha_{u} = 1/\mathcal K$ and $\beta_{u} = \mathcal K$. This temperature controls both the space of models the sampler can reach on a single sweep and how quickly it does so.  For instance, letting $\mathcal K \to \infty$ fixes all the $u_{t}$'s to zero, which results in an intractable infinite sum in equation \eqref{eq:forward-var} corresponding to a full forward-backward pass ignoring the complexity-controlling slice variables. Setting $\mathcal K = 1$ recovers the uniform distribution. Values of $\mathcal K$ tending towards $0$ send all the $u_{t}$'s to 1, which limits the computational complexity of the sampler (fewer paths are explored on each sweep), but causes the chain to mix more slowly. Because of the deterministic state transitions introduced by using a remaining duration counter, sampling with a temperature greater than one was generally found to be beneficial in terms of faster mixing.

{\bf Sampling $\bsym\gamma$, $\bsym\pi$, and $\bsym w$} Beam sampling $\bsym z$ results in a sequence of ``observed'' transitions.  Given these (and respective priors), $\bsym\gamma$ and $\bsym\pi$ are sampled.  To do this, the observed transitions are sequentially seated in a Chinese restaurant franchise (CRF) \cite{Teh:2006b} to instantiate  counts for the dependent DPs.  That is, with the observed counts from $\tilde D_{m}$ contained in the vector $\bsym C_{m}$ (i.e.~$C_{mk}$ is the number of observed transitions from state $m$ to state $k$), we denote the number of tables serving $\theta_{k}$ in the restaurant representation of $\tilde D_{m}$ by $l_{mk}$.  These numbers are equal to the number of draws of $\theta_{k}$ from $D_{m}$. A Gibbs sample of the $\bsym l$'s can be generated by running a Chinese restaurant process, with customers $\bsym C_{m}$ and keeping track of the number of tables generated. That is, to sample $l_{mk}$ we seat each of the $C_{mk}$ customer one at a time. The probability of that customer sitting at a new table is proportional to $\alpha_{1}\beta_{mk}$ while the probability of sitting at an existing table is simply the proportional to the number of customers already seated (we don't care which of the existing tables the new customer sits at). Thus, the process for generating $l_{mk}$ is
\eqan
l_{mk}^{(1)} &=& 1 \\
l_{mk}^{(j+1)} &=& l_{mk}^{(j)} + \mathbb I \left (X_{j} <  \frac{\alpha_{1}\beta_{mk}}{(\alpha_{1}\beta_{mk} + j)} \right) \\
l_{mk} &=& l_{mk}^{(C_{mk})} 
\enan
with $X_{j} \sim \mathsf{Uniform}(0, 1)$.

In order to sample the weights $\bsym \gamma$ of the independent gamma processes, we follow the auxiliary variable method of Rao \& Teh (2009) \cite{Rao:2009}, which we re-derive for the case of the \SIHMM here. We are interested in the posteriors of $\{\gamma_{m}\}_{m \in \mathcal T}$ and $\gamma_{+}$ in the IED-HMM (equations \eqref{eq:ied-gamma-m} and \eqref{eq:ied-gamma-plus}) or $\{\gamma_{m}\}_{m \in \mathcal T}$ and $\{\gamma_{m+}\}_{m \in \mathcal T}$ in the ILR-HMM  (equations \eqref{eq:ilr-gamma-m} and \eqref{eq:ilr-gamma-m-plus}). Recall that $\mathcal A$ is a partition of $\mathbb N$ that allows for the reconstruction of the restricting sets $V_{m}$ and $\mathcal A_{m}$ is a subset of sets in $\mathcal A$, the union of which is $V_{m}$. We can generically represent $\bsym \gamma$ by the set $\{\gamma_{R}\}_{R \in \mathcal A}$, where (cf.~equation \eqref{eq:DPmixture})
\eqn \gamma_{R} \sim \GammaRV\left(\alpha_{0}\textstyle\sum_{k \in R} w_{k}, 1\right). \enn

Let dots in subscripts indicate summation over that index and define $l_{\cdot R} := \sum_{m \in R} l_{\cdot m}$. The the posterior distribution of the collection $\{\gamma_{R}\}_{R \in \mathcal A}$ is 
\begin{align} 
p(&\{\gamma_{R}\}_{R \in \mathcal A} |\bsym l, \bsym w) = \nonumber \\
&\left(\textstyle\prod_{R \in \mathcal A} \gamma_{R}^{\mu_{R}(\Theta) + l_{\cdot R} - 1}e^{-\gamma_{R}}\right) \textstyle\prod_{j}\left(\textstyle\sum_{R \in \mathcal A_{m}} \gamma_{R}\right)^{-l_{j\cdot}}. \label{eq:gamma-post-prob}
\end{align}
To make sampling tractable, we introduce the auxiliary variables $\bsym L = \{ L_{j} \}_{j \in \mathcal T}$ and use the Gamma identity
\eqn \Gamma(l_{j\cdot})\left(\textstyle\sum_{R \in \mathcal A_{m}} \gamma_{R}\right)^{-l_{j\cdot}} = \int_{0}^{\infty} L_{j}^{l_{j\cdot}-1}e^{-\sum_{R \in \mathcal A_{j}}\gamma_{R}L_{j}}dL_{j} \enn
which combined with \eqref{eq:gamma-post-prob} implies that the joint posterior probability of $\{\gamma_{R}\}$ and $\bsym L$ is 
\begin{align}  
p(&\{\gamma_{R}\}_{R \in \mathcal A}, \bsym L |\bsym l, \bsym w) \propto  \\
&\left(\textstyle\prod_{R \in \mathcal A} \gamma_{R}^{\mu_{R}(\Theta) + l_{l_{\cdot R}} - 1}e^{-\gamma_{R}}\right)\left(\textstyle\prod_{j} L_{j}^{l_{j\cdot}-1}e^{-\sum_{R \in \mathcal A_{j}}\gamma_{R}L_{j}} \right) \nonumber. 
\end{align}
Therefore, the $\gamma_{R}$'s and $L_{i}$'s can be Gibbs sampled according to 
\eqan
\gamma_{R} &\sim& \GammaRV\left(\mu_{R}(\Theta) + l_{\cdot R}, \left(1 + \textstyle\sum_{j \in J_{R}} L_{j}\right)^{-1}\right)\quad \\
L_{j}  &\sim& \GammaRV\left(l_{j\cdot}, \left(\textstyle\sum_{R \in \mathcal A_{j}} \gamma_{R}\right)^{-1}\right),
\enan
where $J_{R} = \{ j | R \in \mathcal A_{j} \}$. Noting that $\mu_{R}(\Theta) = \alpha_{0}\sum_{k \in R} w_{k}$, we now consider the two concrete cases of $\bsym\gamma$. For the IED-HMM
\begin{align}
\gamma_{m} &\sim \mathsf{Gamma}\bigg(l_{\cdot m} + \alpha_{0}w_{m},  \left(1 + L_{\setminus m}\right)^{-1}\bigg)  \\
\gamma_{+} &\sim \mathsf{Gamma}\left(\alpha_{0}w_{+}, \left(1 + \textstyle\sum_{j \in \mathcal T} L_{j}\right)^{-1}\right) \\
L_{j} &\sim \mathsf{Gamma}\left(l_{j\cdot}, \left(\gamma_{+} + \textstyle\sum_{j \in \mathcal T \setminus \{i\}} \gamma_{i}\right)^{-1}\right),
\end{align}
where $L_{\setminus m} = \textstyle\sum_{j \ne m}^{M} L_{j}$ and for the ILR-HMM
\begin{align}
\gamma_{m} &\sim \mathsf{Gamma}\bigg(l_{\cdot m} + \alpha_{0}w_{m},  \left(1 + L_{>m} \right)^{-1}\bigg)  \\
\gamma_{m+} &\sim \mathsf{Gamma}\left(\alpha_{0}\textstyle\sum_{k \in R_{m+}}w_{k}, \left(1 + L_{>m}\right)^{-1}\right) \\
L_{j} &\sim \mathsf{Gamma}\left(l_{j\cdot}, \left(\gamma_{+} + \textstyle\sum_{i \ne j}^{M} \gamma_{i}\right)^{-1}\right)
\end{align}
where $L_{>m} = \sum_{j > m} L_{j}$.

\begin{figure*}[t]
\centering
\includegraphics[width=.9\textwidth,trim= 0cm 1cm 0cm 1cm]{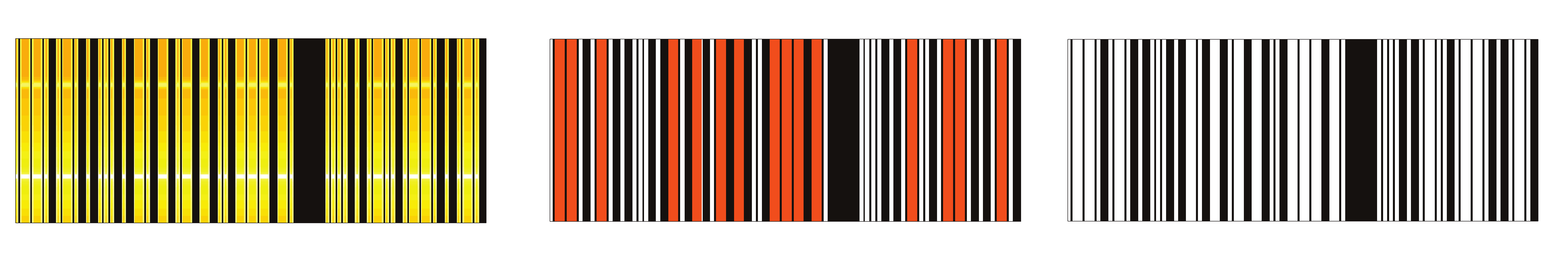}
\caption{IED-HMM segmentation of morse code. {\em Left:}  Spectrogram of Morse code audio.  {\em Middle:}  IED-HMM MAP state sequence {\em Right:} sticky HDP-HMM MAP state sequence. The IED-HMM correctly learns three states (off -- black, dot -- white, and dash -- orange) even though the dot and dash states can only be distinguished by duration.  Non-explicit-duration HMMs like the sticky HDP-HMM are only able to learn two because of this.}
\label{fig:morse_code_spectrogram}
\end{figure*}

After sampling $\bsym\gamma$, the matrix $\bsym\beta$ is calculated deterministically using equation \eqref{eq:beta}. The stick weights $\bsym w$ were sampled using Metropolis Hastings updates and
the rows of the transition matrix $\bsym\pi$ are sampled according to
\eqn \bsym \pi_{m} \sim \mathsf{Dirichlet}(\bsym C_{m} + \alpha_{1}\bsym\beta_{m}). \enn 
{\bf Sampling $\bsym\theta$, $\bsym\lambda$, and hyperparameters} Sampling of $\bsym\theta$ and $\bsym\lambda$ depends on the choice of prior distributions $H_{\Theta}$ and $H_{r}$, and data distributions $F_{\Theta}$ and $F_{r}$.  For standard choices, straightforward MCMC sampling techniques can be employed. The concentration parameters $\alpha_{0}$ and $\alpha_{1}$ are sampled via Metropolis-Hastings. 


\subsubsection{Considerations During Forward Inference}
\label{sec:forward_consideration}
In a non-parametric model, not all the infinite parameters (specifically, states) can be explicitly represented. Since an unknown number of the parameters will be needed, in the case of infinite HMMs there must be a procedure to instantiate state-specific parameters as new states are needed during inference. This instantiation takes place during forward inference, the details of which are described below in the context of the IED-HMM. However, the same principles apply to the ILR-HMM and other parameterizations of the \SIHMM. 

Recall that in the IED-HMM, $\pi_{m+}$ is the total probability of transitioning from state $m$ to some unused state. When calculating the forward variables $\alpha_{t}(z_{t})$, if $\pi_{m+} > u_{t}$ for some $m$ and $t$, then it is possible for a transition into one of the merged states to occur. In this case, merged states
must be instantiated on the fly until all the $\pi_{m+}$ are small enough that the beam sampler will not consider them during the forward filtering step: for all $m$ and $t$, we must ensure that $\pi_{m+} < u_{t}$ . 

To instantiate a merged state $M \notin \mathcal T$, note that $\gamma_{+}$ is the total weight for all unobserved states, i.e.~the total weight for the draw $G_{+}$ from the gamma process $\Gamma \textnormal P(\mu_{R_{+}})$ (cf.~\eqref{eq:DPmixture} and \eqref{eq:ied-gamma-plus}). Thus, state $M$ must have weight $\gamma_{M} < \gamma_{+}$. Since the normalized weight $\gamma_{M}/\gamma_{+}$ is a weight from a DP, it can be sampled using the stick breaking construction
\eqan
b_{M} &\sim& \mathsf{Beta}(1, \mu_{+}(\Theta)) \\
\gamma_{M} &=& b_{M}\gamma_{+} \\
\gamma_{+} &\leftarrow& (1-b_{M})\gamma_{+}.
\enan
The normalization terms for the $\bsym\beta_{m}$'s do not change since the total weight of accessible states from state $m$ remains constant. But $\beta_{mM}$ (and thus $\pi_{mM}$) must be instantiated and $\beta_{m+}$ (and thus $\pi_{m+}$) must be updated. The updates to $\bsym\pi$ can be accomplished by noting that if $\mathcal T = \{ v_{1},\dots,v_{K}\}$, then $\pi_{mM}$ and $\pi_{m+}$ are two components of a draw from $\mathsf{Dirichlet}(\alpha_{1}\beta_{mv_{1}},\dots,\alpha_{1}\beta_{mv_{K}}, \alpha_{1}\beta_{mM+1}, \alpha_{1}\beta_{m+})$, so a draw from $\mathsf{Dirichlet}(\alpha_{1}\beta_{mM+1}, \alpha_{1}\beta_{m+})$ (i.e.~the beta distribution) gives the proportion of the old $\pi_{m+}$ that stays on $\pi_{m+}$ and the proportion that is broken off to form $\pi_{mM}$
\eqan
b_{mM} &\sim& \mathsf{Beta}(\alpha_{1}\beta_{mM}, \alpha_{1}\beta_{m+}) \\
\pi_{mM+1} &=& b_{mM}\pi_{m+} \\
\pi_{m+} &\leftarrow& (1- b_{mM})\pi_{m+}.
\enan
Finally, ${\bsym\pi}_{M}$ is sampled according to \eqref{eq:pi}. 

This state-splitting procedure is repeated as many times as is necessary to ensure that $\pi_{m+} < u_{t}$ for all $m$ and $t$.  Also, it should be noted that this procedure allows for incremental inference in models belonging to the \SIHMM family.

\begin{figure}[t]
\centering
\includegraphics[trim = 2cm 9cm 2cm 9cm, width=.42\textwidth]{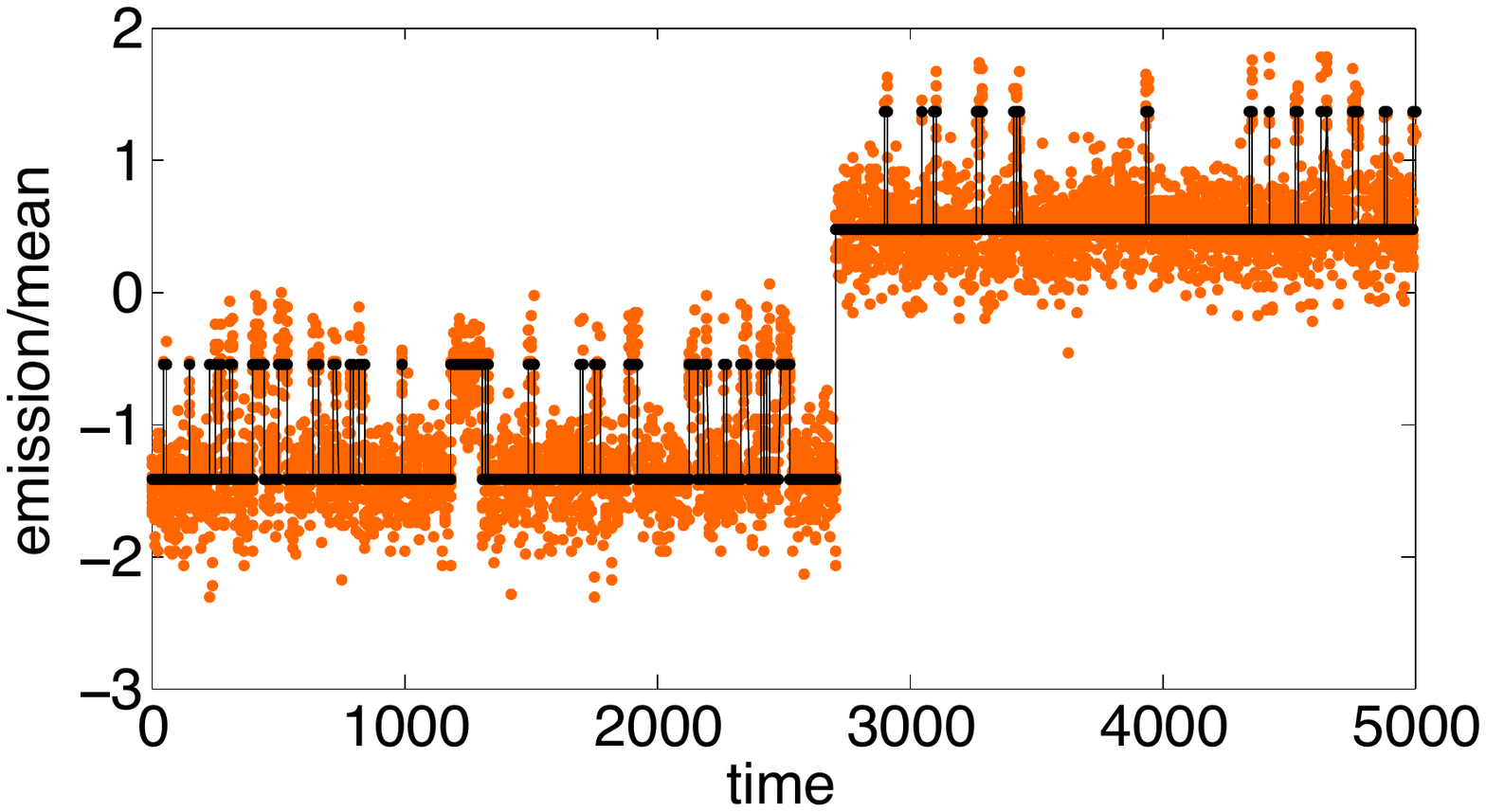} 
\includegraphics[trim = 2cm 9cm 2cm 9cm,width=.42\textwidth]{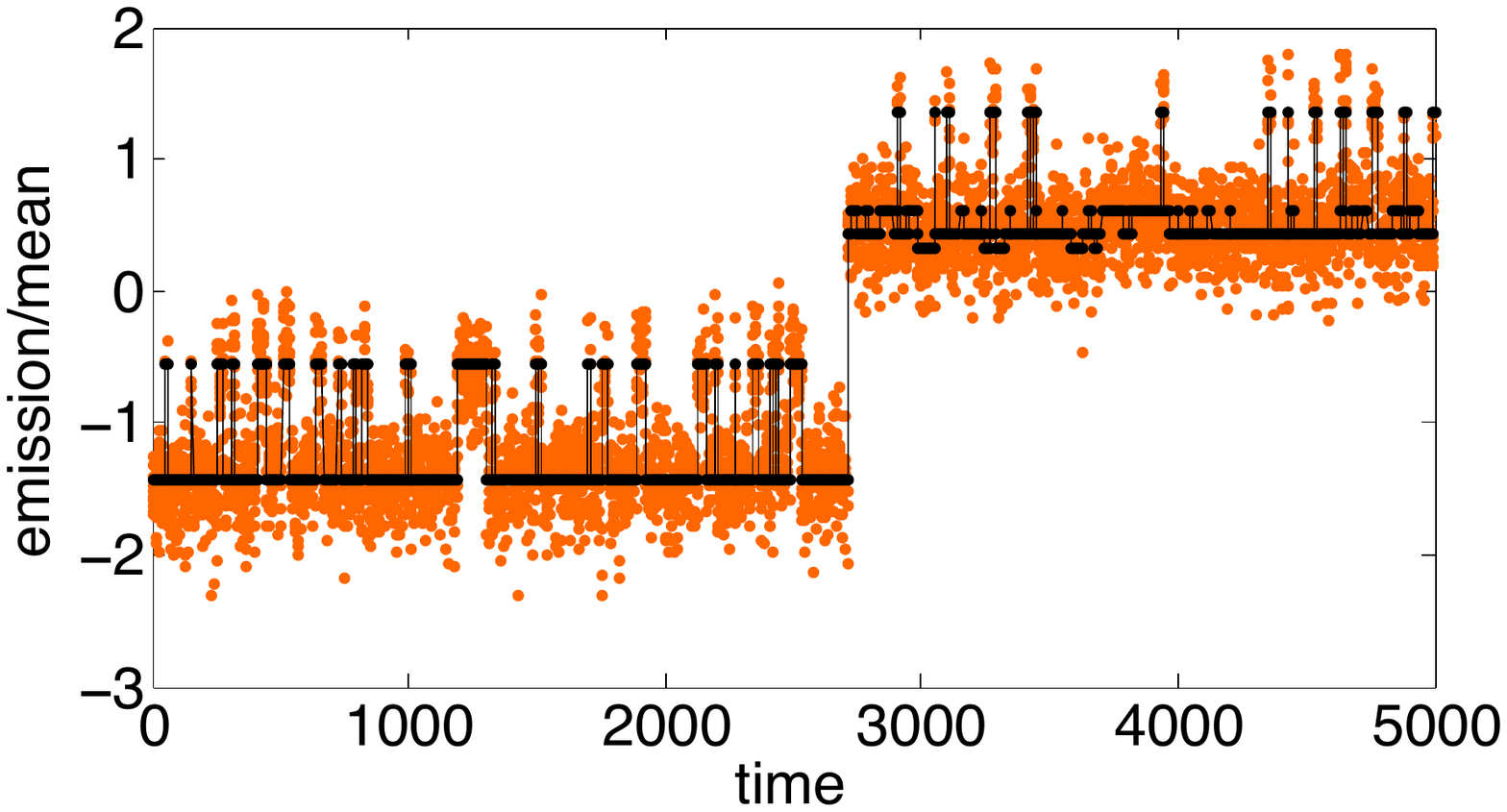}
\caption{Rescaled random telegraph noise data (orange) and MAP estimate of the IED-HMM (black, left) and sticky HDP-HMM (black, right).  Without state-specific duration distribution parameterization even a well-tuned sticky HDP-HMM tends to over-segment.}
\label{fig:rts}
\end{figure}

\section{Experiments}

Infinite HMMs with enhanced state persistence have characteristics that suggest that they will be useful in a number of applied settings.  In this section we show results from a number of simple experiments that illustrate how these models work,  validate the correctness of our novel construction, and highlight some small but important differences between them.

\label{sec:experiments}

\subsection{IED-HMM}

We first illustrate IED-HMM learning on synthetic data. Five hundred datapoints were generated using a 4 state HSMM with Poisson duration distributions (rates $\bsym \lambda = (10, 20, 3, 7)$) and Gaussian emission distributions  (means $\bsym \mu = (-6, -2, 2, 6)$, all unit variance). 
For inference, the emission and duration distributions were given broad priors. The emission distributions were given Normal-scaled Inverse Gamma priors with $\mu_{0} = 0, \nu_{0} = .25, \alpha = 1$, and parameters for the Poisson duration distributions were given $\mathsf{Gamma}(1,10^{3})$ priors. The temperature was set to $\mathcal K = 3$. 

One thousand samples were collected after a burn-in of 100 iterations.  A short burn-in was possible since, due to the forward-backward slice sampler, the Markov chain mixes quite quickly. This can be seen from the autocorrelations of the means and duration distributions associated with the states at fixed times and the joint log-likelihood (see Figure \ref{fig:edihmm-autocorr}). Figure \ref{fig:overlaid-means-and-state-distribution} shows the highest scoring sample and (inset) a histogram of the state cardinalities of the models explored (82\% of the samples had 4 states)
The inferred posterior distribution of the duration distribution rates and means of the state observation distributions are shown in Fig.~\ref{fig:posteriors-synthetic}.  All contain the true values in regions of high posterior confidence.
\subsubsection{Morse Code}
Morse code consists of a sequence of short and long ``on'' tones.  The frequency spectrum of a sequence of Morse code audio (8KHz., 9.46 sec.)  is shown in Fig.~\ref{fig:morse_code_spectrogram}.  Following \cite{Johnson:2010}, we segmented it to illustrate both the utility of explicitly parameterizing duration distributions and to illustrate the correctness of our \SIHMM construction and sampling algorithms.  Figure~\ref{fig:morse_code_spectrogram} also shows that, because each state has its own delayed geometric duration distribution  $ F_r(\lambda_{s_t}) = \mathsf{Geom}(q_{s_t})+d_{s_t}$ (where $\lambda_m = (d_m, q_m), d_m \in \{0, \ldots, 30 \}, q_m \in \mathbb{Z}^+$, and $H_r(d, q) = \frac{1}{30}$), the IED-HMM is able to distinguish short and long tones and assign a unique state identifier to each (using a Gaussian emission model for the first  Mel-frequency Cepstrum coefficient).  This result replicates the results for the same experiment in  \cite{Johnson:2010} using the HDP-HSMM.  Non-explicit-duration HMMs such as the sticky HDP-HMM \cite{Fox:2010tg} can only infer the existence of two states because they cannot distinguish states by duration. 
\subsubsection{Nanoscale Transistor Noise} 
Random telegraph noise (RTN) is the name given to instantaneous temporal changes in modern-day nanoscale transistor current-driving characteristics due to quantum tunneling of charge carriers in and out of potential traps in the device oxide. In macro-scale electronic systems RTN can manifest itself as anything from an annoying flicker of a pixel to complete failure of the entire system. In order to quantify and mitigate the negative effects of RTN, the statistical characteristics of RTN must be well understood \cite{Realov2010}. IED-HMMs are well suited for this task since the duration of the temporal changes is random and of interest and the number of ``error states'' is not known a priori. Figure~\ref{fig:rts} shows the results of using the IED-HMM to a model RTN data in which the domain experts believe that there are four latent states with characteristic duration distributions.  We find that the IED-HMM is able to learn a model in good correspondence with scientist expectation. 

Somewhat surprisingly, the sticky HDP-HMM did not fit the data as well as the IED-HMM. This appears to be because of the shared $\kappa$ prior across all states. The mean duration of states in the IED-HMM MAP sample, which we treat as a proxy for the truth, was approximately 48. In the sticky HDP-HMM MAP sample, the mean duration was only slightly lower, at 40, while the hyperparameters of the MAP sample, in particular $\rho = \kappa/(\alpha_{1} + \kappa)$, gave an expected duration of approximately $1/(1-\rho) = 45$. The single $\rho$ shared across all states gave the sticky HDP-HMM less flexibility to represent a wide variety of expected durations for different states. In the RTS data, one state (with mean of about .5) had long typical durations of between 300 and 400 time steps.  We believe that the sticky HDP-HMM with a single $\kappa$ parameter was biased by the states with means around -.5 and 1.5 that have mean durations of around 10 time steps and transitions for which accordingly occur much more often in the data. This shortcoming could potentially be alleviated by introducing some number of state-specific $\kappa$ values.

\begin{figure*}[tbp]
\centering
\includegraphics[width=.49\textwidth]{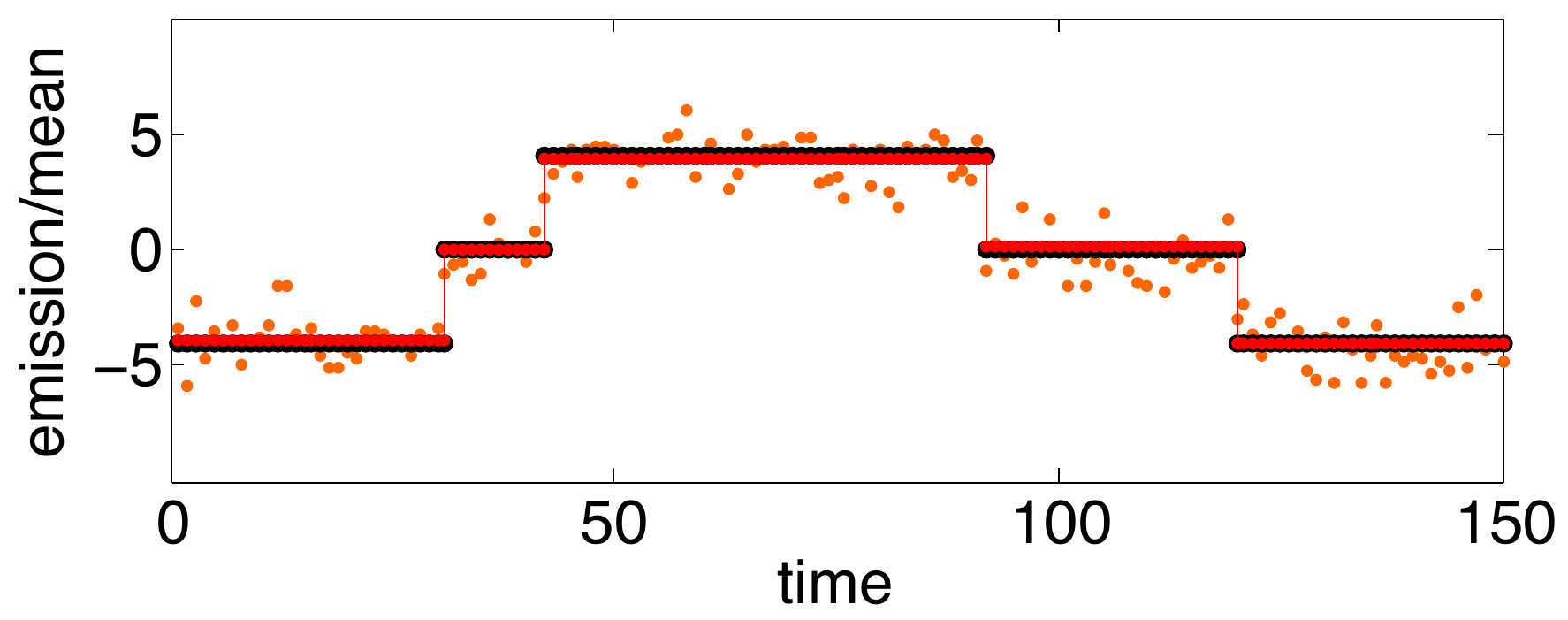} 
\includegraphics[width=.49\textwidth]{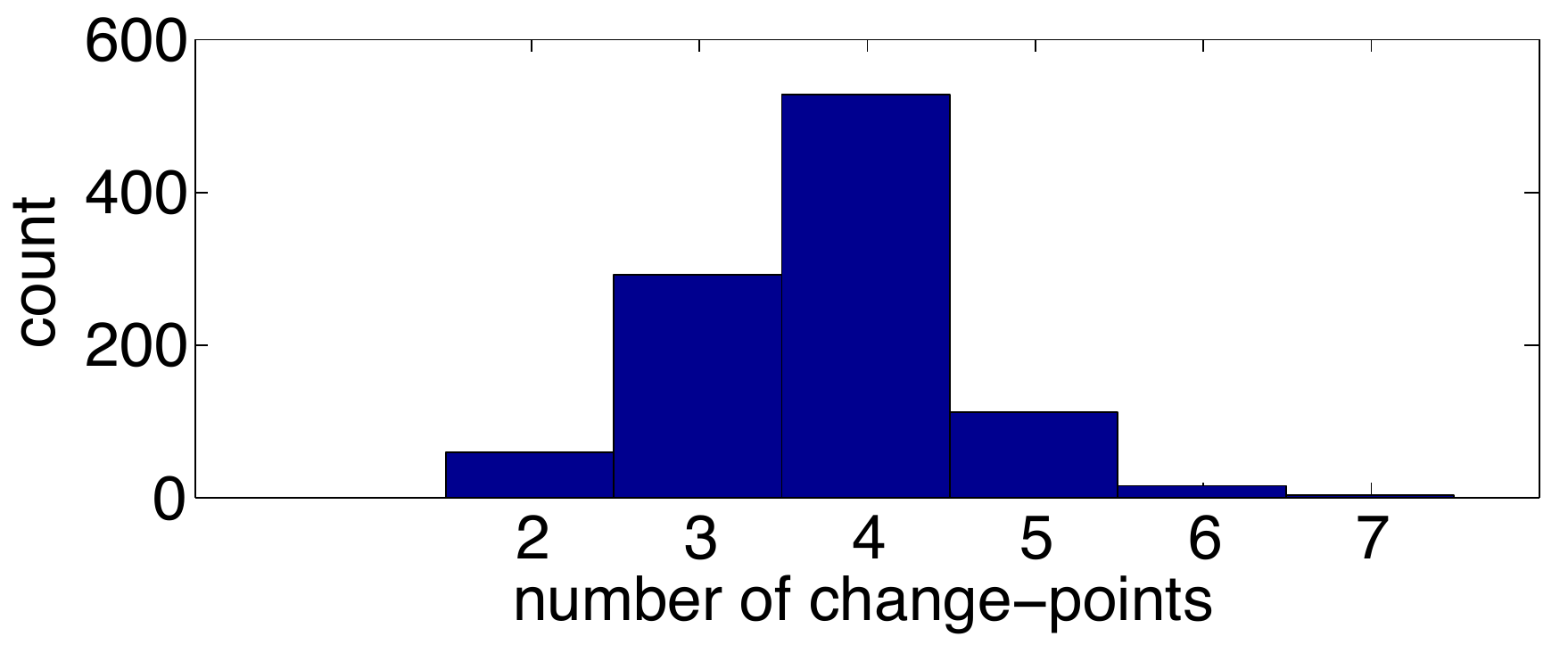} 
\\
\includegraphics[width=.49\textwidth]{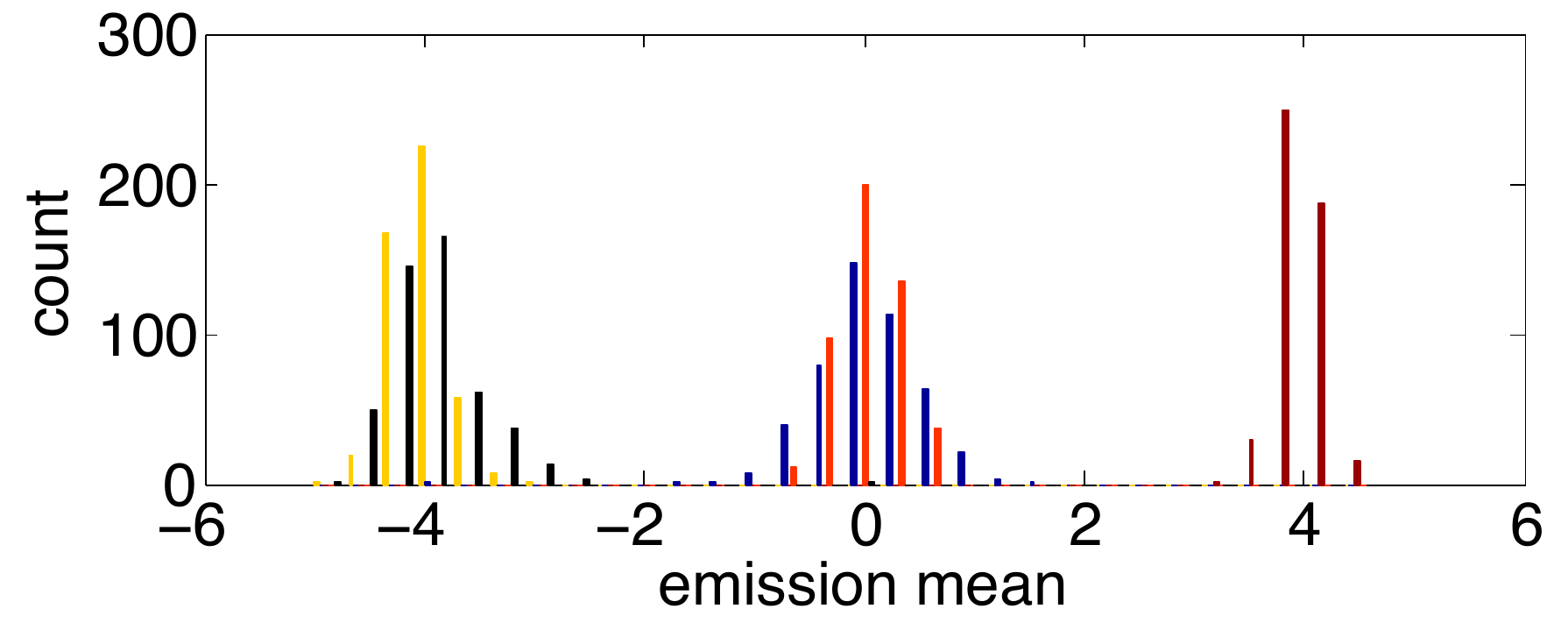}
\includegraphics[width=.49\textwidth]{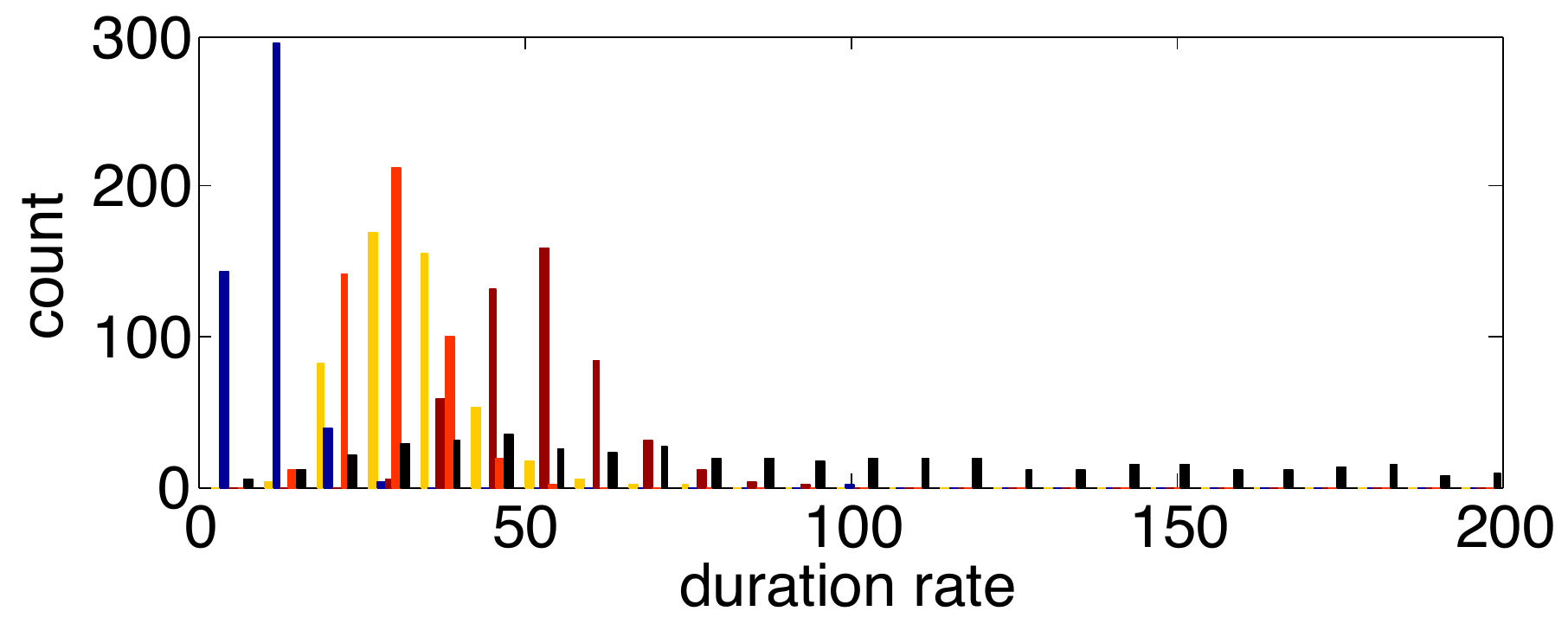}
\caption{{\bf Top left:} Synthetic data generated from a 5 state left-to-right HMM with MAP ILR-HMM sample overlaid; means of the sampled states (red) are overlaid on the true data (orange) and true means (black). {\bf Top right:} Posterior distribution of ILR-HMM change points.  {\bf Bottom:} Posterior distributions of latent state parameters for generated data. 
The final duration is not well defined, hence the posterior has large variance.
}
\label{fig:lr-posteriors-synthetic}
\end{figure*}

\subsection{ILR-HMM}

We illustrate ILR-HMM learning on synthetic data. One hundred and fifty datapoints were generated using a 5 state left-to-right HMM with Poisson duration distributions (rates $\bsym \lambda = (30, 10, 50, 30, -)$) and Gaussian emission distributions  (means $\bsym \mu = (-4, 0, 4, 0, -4)$, all unit variance). The last duration rate is undefined since there is no transition out of the 5th state. Hyperparameters were initialized in the same way as in the IED-HMM synthetic data experiment. 

One thousand samples were collected after a burn-in of 100 iterations.  Figure \ref{fig:lr-posteriors-synthetic} shows the highest scoring sample and a histogram of the number of inferred change-points in the data.
The inferred posterior distribution of the duration distribution rates and means of the state observation distributions are also shown in Fig.~\ref{fig:lr-posteriors-synthetic}.  All contain the true values in regions of high posterior confidence. The posterior duration rate for the final state has large variance, which is consistent with the fact that its duration is not well defined. 

Since each duration is observed only once, it seems possible that the prior over duration distributions could have a strong influence on the ILR-HMM posterior. Therefore, a wide variety of $\alpha_{d}$ values were tested for the $\mathsf{Gamma}(1, \alpha_{d})$ duration rate prior in order to investigate sensitivity of model to the hyperparameters. We found qualitatively similar results for a range reasonable $\alpha_{d}$ values, as small as 20 and as large as 200. 

\subsubsection{Coal Mining Disasters}
\begin{figure}[tbp]
\centering
\includegraphics[width=.5\textwidth]{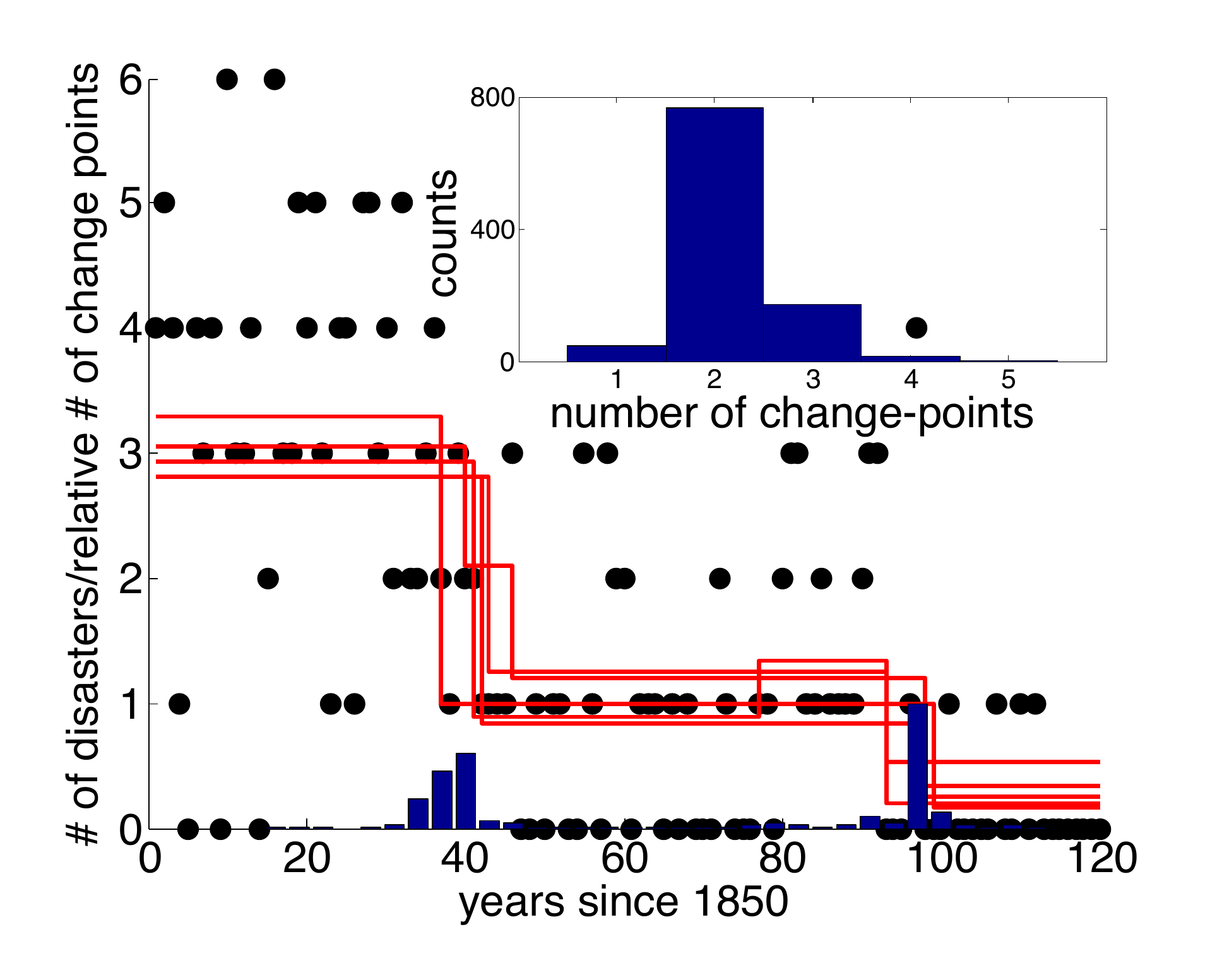}
\caption{Results from using the ILR-HMM to model coal mining disaster data. {\bf Main:} Number of disasters (black dots) and five inferred posterior sample paths (red). A sample path is the mean of the current state. Steps occur at inferred change-points.  The histogram on the horizontal axis shows the relative number of change points inferred to have occurred at each time. {\bf Inset:} Marginal posterior change-point count sample distribution.}
\label{fig:lr-ihmm-coal}
\end{figure}

A well-studied change-point dataset is the number of major coal mining disaster in Britain between 1851 and 1962 \cite{Jarrett:1979}. In previous analyses, such as that by Chib \cite{Chib:1998}, either one or two change-points (i.e.~two or three states) were assumed. We used the ILR-HMM with Poisson emissions and durations (with gamma priors) to model the coal mining data. This allowed us to make no assumptions about the number of change-points. Using 1000 samples the model found two change points with high probability; however the model mixed over multiple interpretations of the data, considering anywhere from one to five change-points (Fig.~\ref{fig:lr-ihmm-coal}, inset). Figure \ref{fig:lr-ihmm-coal} shows the coal mining data and a representative set of posterior samples from the model. The locations of the change-points are well concentrated around 40 and 100 years. This is consistent with previous findings \cite{Green:1995}.

\section{Discussion}

The \SIHMM is a new framework that parameterizes a large number of structured parametric and nonparametric Bayesian HMMs.   It is closely related to the sticky HDP-HMM and the HDP-HSMM but allows direct generalization to infinite HMMs with structured transitions such as the infinite left-to-right HMM.  Inference in the \SIHMM is straightforward and, because of the mathematical particulars of our construction, can largely follow existing sampling techniques for infinite HMMs.  

All of the state-persistence-encouraging infinite HMM constructions solve the same set of problems with only minor differences.  The practical advantages accruing from these enhancements include avoiding the state cardinality selection problem through sampling and control over inferred segmentations through priors that encourage state-persistence. 

There are Bayesian nonparametric equivalents to other popular parametric HMMs including hierarchical \cite{Heller:2009} and factorial \cite{VanGael:2008b} models.  Combining the models reviewed in this paper with those is an area of research ripe for exploration.  In addition, practical application of infinite persistent state HMMs will doubtlessly increase demand for approximate inference approaches suitable for computationally efficient inference.


%
\printbibliography

\end{document}

%% file: my-latex-defs.tex
\long\def\comment#1{}

\newcommand{\eq}{\begin{equation*}}
\newcommand{\en}{\end{equation*}}
\newcommand{\eqa}{\begin{eqnarray*}}
\newcommand{\ena}{\end{eqnarray*}}
\newcommand{\eqn}{\begin{equation}}
\newcommand{\enn}{\end{equation}}
\newcommand{\eqan}{\begin{eqnarray}}
\newcommand{\enan}{\end{eqnarray}}

\newcommand{\pmat}{\begin{pmatrix}}
\newcommand{\pman}{\end{pmatrix}}
\newcommand{\bitems}{\begin{itemize}}
\newcommand{\eitems}{\end{itemize}}
\newcommand{\benum}{\begin{enumerate}}
\newcommand{\eenum}{\end{enumerate}}
\newcommand{\bprf}{\begin{proof}}
\newcommand{\eprf}{\end{proof}}